\documentclass{article}
\usepackage[margin=.8in,left=.8in]{geometry}

\usepackage{amsthm}
\usepackage{amssymb}
\usepackage{amsmath}
\usepackage{mathtools}
\usepackage{adjustbox}
\usepackage{txfonts}
\usepackage{xcolor}
\usepackage{stmaryrd}    
\usepackage{bbding} 
\usepackage{rotating}
\usepackage{xfrac}
\usepackage{enumitem}\setlist{nosep}
\usepackage{ wasysym } 
\usepackage{multirow}
\usepackage{multicol}
\usepackage{hhline}
\usepackage{bbm}

\usepackage[new]{old-arrows}   

\usepackage{mathpartir}

\usepackage{tikz}
\usetikzlibrary{
  backgrounds,
  arrows,
  decorations.pathmorphing,
  decorations.markings,
}
\usetikzlibrary{cd}

\let\PLAINthebibliography\thebibliography
\renewcommand\thebibliography[1]{
  \PLAINthebibliography{#1}
  \setlength{\parskip}{1pt}
  \setlength{\itemsep}{1pt plus .3ex}
}

\usepackage{mathrsfs} 
\DeclareMathAlphabet{\mathpzc}{OT1}{pzc}{m}{it} 

\usepackage[safe]{tipa} 

\usepackage{hyperref}
\hypersetup{
  colorlinks = true,
  allcolors=.
}

\usepackage{cleveref}
\crefformat{section}{\S#2#1#3} 
\crefformat{subsection}{\S#2#1#3}
\crefformat{subsubsection}{\S#2#1#3}

\definecolor{darkblue}{rgb}{0.05,0.25,0.65}
\definecolor{darkgreen}{RGB}{20,140,10}
\definecolor{lightgray}{rgb}{0.9,0.9,0.9}
\definecolor{darkorange}{RGB}{200,100,5}
\definecolor{darkyellow}{rgb}{.91,.91,0}
\definecolor{lightolive}{RGB}{189,183,107}

\theoremstyle{definition}

\def\termscale{.83}

\newcommand{\term}[1]{\scalebox{\termscale}{$#1$}}

\newcommand{\Sets}{
  \mathrm{Set}
}
\newcommand{\Set}{\Sets}

\newcommand{\Groupoids}{\mathrm{Grpd}}

\newcommand{\Categories}{\mathrm{Cat}}

\newcommand{\Actions}[1]{
  {#1}\,\mathrm{Act}
}

\newcommand{\HomotopyQuotient}[2]{{#1}\!\sslash\!{#2}}

\newcommand{\RealSets}
{\ZTwo\Sets}

\newcommand{\ARealSet}[1]{\acts{#1}}

\newcommand{\smooth}[1]{{#1}_{\!\mathrm{sm}}}

\newcommand{\involution}{C}

\newcommand{\ComplexConjugate}[1]{\overline{#1}}

\tikzset{
    vertlabl/.style={anchor=south, rotate=90, inner sep=.5mm}
}

\newcommand{\CyclicGroup}[1]{{\mathbb{Z}_{#1}}}
\newcommand{\ZTwo}{\CyclicGroup{2}}

\def\acts{\raisebox{1.4pt}{\;\rotatebox[origin=c]{90}{$\curvearrowright$}}\hspace{.5pt}}

\newcommand{\Integers}{\mathbb{Z}}

\newcommand{\RealNumbers}{\mathbb{R}}

\newcommand{\RealComplexStructure}{\mathrm{I}}

\newcommand{\RRealNumbers}{\acts\, \ComplexNumbers}

\newcommand{\ImaginaryUnit}{
  \mathrm{i}
}
\newcommand{\ComplexNumbers}{\mathbb{C}}

\newcommand{\Modules}[1]{\mathrm{Mod}_{\scalebox{.7}{$#1$}}}

\newcommand{\DependentModules}[2]{\Modules{#1}^{\scalebox{.7}{$#2$}}}

\newcommand{\defneq}{\equiv}

\newcommand{\braiding}[2]{\mathrm{braid}^{\scalebox{.7}{$#1$}}_{\scalebox{.7}{$#2$}}}

\newcommand{\AnyTypes}{\mathrm{Type}}

\newcommand{\ABundleType}[2]{
\left[
\def\arraystretch{1}\def\arraycolsep{1pt}
  \begin{array}{c}
    {#1}
    \\
    \rotatebox[origin=c]{-90}{$\twoheadrightarrow$}
    \\
    {#2}
  \end{array}
\right]
}

\newcommand{\Types}{\AnyTypes}

\newcommand{\ClassicalTypes}{\mathrm{Cla}\Types}

\newcommand{\LinTypes}{\mathrm{Qu}\AnyTypes}

\newcommand{\QuantumType}{\LinTypes}
\newcommand{\QuantumTypes}{\QuantumType}

\newcommand{\VectorSpace}[1]{#1}

\newcommand{\RealModule}[1]{\mathcal{#1}}

\newcommand{\HilbertSpace}[1]{\mathcal{#1}}

\newcommand{\isa}{:}

\newcommand{\linearly}{\hspace{-2pt}\raisebox{-0pt}{\scalebox{.95}{\rotatebox{30}{$\triangle$}}}\hspace{-0pt}}

\newcommand{\quantized}{\mathrm{Q}}

\newcommand{\classically}{\natural}

\newcommand{\quantumly}{\linearly}

\newcommand{\possibly}{\scalebox{1.06}{$\lozenge$}}

\newcommand{\TensorUnit}{\adjustbox{scale=1.1,raise=-.4pt}{$\mathbbm{1}$}}

\newcommand{\externaltensor}{\otimes}

\newcommand{\heart}{\heartsuit}

\def\LanguageNameScale{1.01}

\newcommand{\HoTT}{\scalebox{\LanguageNameScale}{\tt HoTT}}
\newcommand{\LHoTT}{\scalebox{\LanguageNameScale}{\scalebox{\LanguageNameScale}{\tt L}\HoTT}}

\newcommand{\Quipper}{\scalebox{\LanguageNameScale}{\tt Quipper}}

\begin{document}

\title{Quantum and Reality}

\author{
  Hisham Sati${}^{\ast \dagger}$
  \;\;
  \;\;
  Urs Schreiber${}^{\ast}$
}

\maketitle

\thispagestyle{empty}

\begin{abstract}
 Formalizations of quantum information theory in category theory and type theory, for the design of verifiable quantum programming languages,
 need to express its two fundamental characteristics: (1) parameterized linearity and (2) metricity. The first
 is naturally addressed by dependent-linearly typed languages such as Proto-{\Quipper} or, following our recent observations \cite{Monadology}\cite{EoS}:
 Linear Homotopy Type Theory ({\LHoTT}).
 The second point has received much attention (only) in the form of semantics in ``dagger-categories'', where operator adjoints are
 axiomatized but their specification to Hermitian adjoints still needs to be imposed by hand.

\smallskip
 In this brief note, we describe a natural emergence of Hermiticity which is rooted in principles of equivariant homotopy theory,
 lends itself to homotopically-typed languages and naturally connects
 to topological quantum states classified by twisted equivariant KR-theory.
 Namely, we observe that when the complex numbers are considered as a monoid internal
 to $\ZTwo$-equivariant real linear types, via complex conjugation (the ``Real numbers''), then (finite-dimensional) Hilbert spaces do
 become self-dual
 objects among internally-complex Real modules. This move absorbs the dagger-structure into the type structure;  for instance, a complex
 linear map is unitary iff seen internally to Real modules it is orthogonal.

 \smallskip

  The point is that this construction of Hermitian forms requires of the ambient linear type theory nothing further than a negative unit
  term of tensor unit type.  We observe that just such a term is constructible in plain {\LHoTT}, where it interprets as the non-trivial
  degree=0 element of the $\infty$-group of units of the sphere spectrum, interestingly tying the foundations of quantum theory to homotopy
  theory. We close by indicating how this observation allows for encoding (and verifying) the unitarity of quantum gates and of quantum channels
  in quantum languages embedded into {\LHoTT}, as described in \cite{Monadology}.
\end{abstract}

\vspace{1cm}


\vfill

\hrule
\vspace{5pt}

{
\footnotesize
\noindent
\def\arraystretch{1}
\tabcolsep=0pt
\begin{tabular}{ll}
${}^*$\,
&
Mathematics, Division of Science; and
\\
&\
Center for Quantum and Topological Systems,
\\
&
NYUAD Research Institute,
\\
&
New York University Abu Dhabi, UAE.
\end{tabular}
\hfill
\adjustbox{raise=-15pt}{
\href{https://nyuad.nyu.edu/cqts}{\includegraphics[width=3cm]{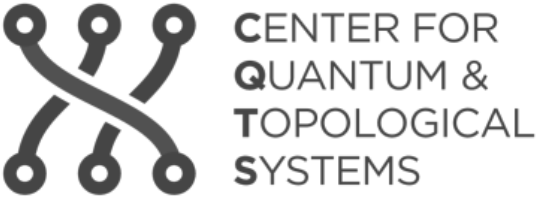}}
}

\vspace{1mm}
\noindent ${}^\dagger$The Courant Institute for Mathematical Sciences, NYU, NY

\vspace{.2cm}

\noindent
The authors acknowledge the support by {\it Tamkeen} under the
{\it NYU Abu Dhabi Research Institute grant} {\tt CG008}.

}

\newpage

Quantum theory rests on two principles: (1) parameterized linearity and (2) metricity. The former is embodied by tensor-linear algebra
and controls quantum phenomena such as the no-cloning/no-deleting and entanglement. The latter is embodied by quadratic
forms and controls (via the Born rule) the probabilistic content of quantum physics and hence its relation to observable reality.
(Extensive review and references may be found in \cite[\S 1.1, 1.2]{Monadology}.)

\vspace{-3mm}
\paragraph{The Hermitian core of Quantum physics.}
Ever since the foundations of quantum physics were laid, \cite[p. 64]{vonNeumann30}\cite[p. 21]{vonNeumann32},
the quadratic form on vector spaces $\VectorSpace{H}$ of quantum states is known to arise from a Hermitian form\footnote{
  For the purpose of this article, ``Hermitian form'', ``metric'',  and ``inner product'' all imply the non-degeneracy and
  (conjugate-)symmetry of the pairing but not necessarily its signature.
}
$\langle \cdot \vert \cdot \rangle$ (e.g. \cite[\S A.1]{Landsman17}),
hence complex {\it sesqui-}linear instead of complex bi-linear (e.g. \cite[\S V.1.1]{Bourbaki81}):
\begin{equation}
  \label{HermitianFormInIntroduction}
  \big\langle
    c_1 \psi_1
    \big\vert
    c_2\psi_2
  \big\rangle
  \;=\;
  \overline{c}_1
  c_2
  \langle
  \psi_1
  \vert
  \psi_2
  \rangle
  \,,
  \;\;\;\;\;\;\;
  \langle \psi_2 \vert \psi_1 \rangle
  \;=\;
  \overline{
    \langle \psi_1 \vert \psi_2 \rangle
  }
  \,.
\end{equation}
The whole probabilistic interpretation of quantum physics via the Born rule (see pointers in \cite[p. 23]{Monadology}), hence its observable content, relies on this mild {\it failure},
if you will, of complex-linearity in an otherwise complex-linear theory.

\smallskip
This is not a logical tautology, but a core feature of quantum physics: Complex bi-linear forms do certainly play a role elsewhere in mathematics,
for instance as Killing forms on semisimple complex Lie algebras. Here we consider the question at the foundations of quantum theory: How to
fundamentally (category-theoretically) think of the phenomenon of Hermitian structure underlying quantum physics?

\medskip

\noindent
{\bf The subtle abstract nature of Hermitian structure.}
Abstractly, sesqui-linear forms are more subtle than complex bi-linear forms, since the plain monoidal category $(\mathcal{C}, \otimes, \TensorUnit)$
of complex vector spaces $(\Modules{\ComplexNumbers}, \otimes_{{}_{\ComplexNumbers}}, \mathbb{C})$ does {\it not} support them: While a bi-linear form
is equivalently an isomorphism in this category to the dual space,
$\begin{tikzcd}[column sep=10pt]\RealModule{V} \ar[r, shorten=-2pt, "\sim"{yshift=-1pt, pos=.4}] & \RealModule{V}^\ast := \mathrm{Map}_{\mathbb{C}}(\RealModule{V}, \mathbb{C})\end{tikzcd}$ (cf. \cite{Selinger12}), a sesquilinear form is  an anti-linear
such isomorphism or equivalently a complex-linear isomorphism but to the anti-dual space $\RealModule{V} \xrightarrow{\,\sim\,} \overline{\RealModule{V}^\ast}$:
The first does not exist in $\mathrm{Mor}(\Modules{\ComplexNumbers})$, while the latter does not have a universal construction among
$\mathrm{Obj}(\Modules{\ComplexNumbers})$ -- not without invoking further structure. \footnote{That further structure may and often
is taken to be that of a {\it dagger-category} (\cite{Selinger07}, cf. \cite[\S 2.3]{HeunenVicary19}, which recovers inner products
according to \cite[Def. 7.5]{AbramskyCoecke04}, see e.g. \cite{HeunenVicary19} for general review and \cite{StehouwerSteinebrunner23} for more recent developments). Just to highlight that
dagger-structure serves to parameterize the issue more than it illuminates it: Besides the usual dagger structure for Hermitian
inner products, $\Modules{\ComplexNumbers}$ also carries a dagger-structure encoding bilinear inner products. Dagger-theory
by itself does not select one over the other for the purpose of quantum theory.}

\medskip

\noindent
{\bf The problem in typed quantum programming languages.}
This does become an issue when formalizing quantum information theory (eg. \cite{NielsenChuang10}), such as in formulating typed functional quantum programming languages: Modern quantum type systems
such as Proto-{\tt Quipper} \cite{FuKishidaSelinger20} or {\LHoTT} \cite{Riley22}\cite{Riley23} (as explained in \cite{Monadology}),
axiomatize {\it linear types} with semantics in (parameterized versions of) distributive symmetric closed monoidal categories \cite{RiosSelinger18}\cite{RileyFinsterLicata21}\cite{EoS}, but do not explicitly
express anti-linear structure. In these formal languages, one can speak about inner products in terms of self-dual objects in any such monoidal category,
but one just cannot speak directly about sesqui-linearity.

\smallskip
Of course, it is possible (see \cite{MyersRiley24}) to add inference rules to linear type theories which force an involution $\overline{(\mbox{-})}$
on the type system such that function types $\HilbertSpace{H}_1 \to \overline{\HilbertSpace{H}}_2$ behave like spaces of antilinear maps.
However, part of the beauty at least of {\LHoTT} is that it simultaneously serves as a proof system for general abstract parameterized stable
homotopy theory (cf. \cite[\S 1.5]{Monadology}); and here we would rather discover sesqui-linear structure than to artificially enforce it.

\medskip

\noindent
{\bf Motivation: Anyonic quantum states as Real K-modules.}
Concretely, we are motivated by our recent observation \cite{Def} \cite{Ord}\cite{TQP} that spaces of anyonic quantum states --- thought to be needed
for topological quantum computation --- have a profound description as (twisted equivariant) {\it cohomology groups} (of configuration spaces of points),
receiving the Chern-character map from the same twisted equivariant K-theory which is thought to classify topological ground states of crystalline
quantum materials. This makes the quantum state spaces appear via modules over the Eilenberg-MacLane spectrum $H^{\mathrm{ev}}\ComplexNumbers$
which is the complex-rationalization of the complex K-theory spectrum $\mathrm{KU}$ (the classifying spectrum for ``K-theory with complex coefficients'').

\smallskip
Or rather --- and this is the pivotal point --- since generally these topological phases are classified by Real K-theory (with capital ``R'')
whose equivariant spectrum $\ZTwo \!\acts \, \mathrm{KU}$ encapsulates the $\ZTwo$-action by complex conjugation (e.g. \cite[Ex. 4.5.4]{Bund}),
anyonic quantum ground states really arise via modules of the equivariant EM-spectrum $\ZTwo \!\acts \, H^{\mathrm{ev}} \ComplexNumbers$,
being modules over the monoid $\ZTwo \!\acts \, \ComplexNumbers$ of complex numbers equipped with their involution by complex-conjugation.

\medskip
\noindent
{\bf Real modules and Real vector bundles.}
This suggests that the category of $\ZTwo \!\acts \, \ComplexNumbers$-modules over base $\ZTwo$-spaces is a good context for understanding Hermitian forms.
This is Atiyah's category of
{\it Real vector bundles} \cite{Atiyah66}. The now traditional capitalization ``Real'' was introduced by later authors
who felt the need for disambiguation, while the original \cite{Atiyah66} just speaks of ``real vector bundles'',
suggestive of the relevance of this generalization. This hunch is maybe further confirmed by our following observations:

\medskip

\noindent
{\bf The Real modules.}
First to recall (from \cite[\S 1]{Atiyah66}) that Real vector bundles over the point (we will say {\it Real modules}, for short) are simply $\ComplexNumbers$-vector spaces $\VectorSpace{V}$ equipped with anti-linear involutions $C$, and homomorphisms between them are $\ComplexNumbers$-linear maps which intertwine these involutions:

\vspace{-3mm}
\begin{equation}
  \label{CategoryOfRealModules}
  \Modules{\RRealNumbers}
  \;\;\;\;\;
  \defneq
  \;\;\;\;\;
  \left\{
  \adjustbox{
    raise=-3pt
  }{
  \begin{tikzcd}[
    column sep=50pt
  ]
    \underset{
      \mathclap{
        \raisebox{-5pt}{
        \scalebox{.7}{
          \color{darkblue}
          \bf
          \def\arraystretch{.8}
          \begin{tabular}{c}
            Real
            \\
            module
          \end{tabular}
        }
        }
      }
    }{
      \VectorSpace{V}_1
    }
    \ar[out=40+90, in=-40+90,
      looseness=3.3,
      shorten <=-3pt,
      shorten >=-3pt,
      "\scalebox{1}{${
          \hspace{3pt}
          \mathclap{
            C_1
          }
          \hspace{4pt}
        }$}"{pos=.5, description}
    ]
    \ar[
      rr,
      "{ f }",
      "{
        \scalebox{.7}{
          \color{darkgreen}
          \bf
          Real homomorphism
        }
      }"{swap, yshift=-1pt}
    ]
    &&
    \underset{
      \mathclap{
        \raisebox{-5pt}{
        \scalebox{.7}{
          \color{darkblue}
          \bf
          \def\arraystretch{.8}
          \begin{tabular}{c}
            Real
            \\
            module
          \end{tabular}
        }
        }
      }
    }{
      \VectorSpace{V}_2
    }
    \ar[out=40+90, in=-40+90,
      looseness=3.3,
      shorten <=-3pt,
      shorten >=-3pt,
      "\scalebox{1}{${
          \hspace{3pt}
          \mathclap{
            C_1
          }
          \hspace{4pt}
        }$}"{pos=.5, description}
    ]
    \\[-15pt]
    \overset{
      \mathclap{
        \raisebox{6pt}{
          \scalebox{.7}{
            \color{darkblue}
            \bf
            \def\arraystretch{.8}
            \begin{tabular}{c}
              $\ComplexNumbers$-vector
              \\
              space
            \end{tabular}
          }
        }
      }
    }{
     \VectorSpace{W}_1
    }
    \ar[
      rr,
      "{ f }",
      "{
        \scalebox{.7}{
          \color{darkgreen}
          \bf
          \scalebox{1.4}{$\ComplexNumbers$}-linear map
        }
      }"{swap}
    ]
    \ar[
      d,
      <->,
      "{ \involution_1 }",
      "{
        \scalebox{.6}{
          \def\arraystretch{.8}
          \begin{tabular}{c}
            anti-lin
            \\
            involut.
          \end{tabular}
        }
      }"{sloped, swap, yshift=-1pt}
    ]
    &&
    \overset{
      \mathclap{
        \raisebox{6pt}{
          \scalebox{.7}{
            \color{darkblue}
            \bf
            \def\arraystretch{.8}
            \begin{tabular}{c}
              $\ComplexNumbers$-vector
              \\
              space
            \end{tabular}
          }
        }
      }
    }{
      \VectorSpace{W}_2
    }
    \ar[
      d,
      <->,
      "{ \involution_2 }"{swap},
      "{
        \scalebox{.6}{
          \def\arraystretch{.8}
          \begin{tabular}{c}
            anti-lin
            \\
            involut.
          \end{tabular}
        }
      }"{sloped, yshift=0pt}
    ]
    \\
    \VectorSpace{W}_1
    \ar[
      rr,
      "{ f }"
    ]
    &&
    \VectorSpace{W}_2
  \end{tikzcd}
  }
  \right\}
  \,.
\end{equation}

\vspace{1mm}
\noindent    This becomes a symmectric closed monoidal category $(\mathcal{C}, \otimes, \TensorUnit, \sigma)$ \cite{EilenbergKelly66} under forming
$\ComplexNumbers$-linear tensor products equipped with the tensor involutions; the tensor unit $\TensorUnit$ is the ``Real numbers'' $\RRealNumbers$
given by $\ComplexNumbers$ equipped with its involution by complex conjugation $\overline{(\mbox{-})}$ and the symmetric braiding $\sigma$ is that
of $\Modules{\ComplexNumbers}$ and swapping the involutions.
This monoidal category of Real modules is
equivalent to that of $\RealNumbers$-vector spaces, via  complexification (cf. \cite[p. 369]{Atiyah66}):
\vspace{-2mm}
\begin{equation}  \label{EquivalenceBetweenRVectorSpaceAndRealVectorBundlesOverPoint}
  \begin{tikzcd}[
    row sep=-3pt,
    column sep=30pt
  ]
    \overset{
     \mathclap{
       \raisebox{5pt}{
       \scalebox{.7}{
         \color{darkblue} \bf
         \scalebox{1}{$\RealNumbers$}-vector spaces
       }
       }
     }
    }{
    \big(
      \Modules{\RealNumbers},
      \otimes_{{}_{\RealNumbers}},
      \RealNumbers,
      \sigma_{\!{}_\RealNumbers}
    \big)
    }
    \ar[
      rr,
      "{
        \sim
      }",
      "{
        \scalebox{.6}{
          \color{darkgreen}
          \bf
          complexification
        }
      }"{swap, yshift=-1pt}
    ]
    &&
    \overset{
     \mathclap{
       \raisebox{3pt}{
       \scalebox{.7}{ \bf
         \color{darkblue}
         Real modules
       }
       }
     }
    }{
    \big(
      \Modules{\RRealNumbers},
      \otimes_{\scalebox{.6}{$\RRealNumbers$}},
      \RRealNumbers,
      \sigma_{{}_\RealNumbers}
    \big)
    }
    \\[10pt]
    \VectorSpace{V}
    \ar[
      d,
      "{ \phi }"{swap}
    ]
    \ar[
      rr,
      shift right=20pt,
      phantom,
      "{ \longmapsto }"
    ]
    &&
    \VectorSpace{V}
      \otimes_{{}_{\RealNumbers}}
      \!
    \ComplexNumbers
    \ar[out=-30, in=30,
      looseness=3.2,
      shorten <=-3pt,
      "\scalebox{1}{${
          \hspace{2pt}
          \mathclap{
            \overline{(\mbox{-})}
            \mathrlap{
              \scalebox{.6}{
                \color{gray}
                complex conjugation
              }
            }
          }
          \hspace{4pt}
        }$}"{pos=.5, description}
    ]
    \ar[
      d,
      "{
        \phi
        \,
        \otimes_{{}_{\RealNumbers}}
        \ComplexNumbers
      }"
      {swap}
    ]
    \\[30pt]
    \VectorSpace{V}'
    &&
    \VectorSpace{V}
      \otimes_{{}_{\RealNumbers}}
      \!
    \ComplexNumbers
    \ar[out=-30, in=30,
      looseness=3.2,
      shorten <=-3pt,
      "\scalebox{1}{${
          \hspace{2pt}
          \mathclap{
            \overline{(\mbox{-})}
          }
          \hspace{4pt}
        }$}"{pos=.5, description}
    ]
  \end{tikzcd}
\end{equation}
\vspace{-.5cm}

\noindent
We observe that it is interesting to pass inner product spaces through this equivalence:

\smallskip
\noindent
{\bf Hermitian forms are Real inner products.}
It is classical that $\RealNumbers$-inner product spaces $(\VectorSpace{V}, g)$ with isometric complex structure $J$ (e.g. \cite[Def. 1.2.1, 1.2.11]{Huybrechts04})

\vspace{-.4cm}
\begin{equation}
  \label{RealPartOfHermitianForm}
  \def\arraystretch{1.3}
  \VectorSpace{V}
  \,\in\,
  \Modules{\RealNumbers}
  \,,
  \hspace{.4cm}
  \begin{array}{l}
    J \,:\,
    \VectorSpace{V}
    \xrightarrow{\;}
    \VectorSpace{V}
    \,,
    \\
    g
      \,:\,
    \VectorSpace{V}
    \otimes_{{}_{\RealNumbers}}
    \VectorSpace{V}
    \xrightarrow{\;}
    \RealNumbers
    \,,
  \end{array}
  \hspace{.5cm}
  \begin{array}{l}
  J \circ J \,=\, - \mathrm{id}_{\VectorSpace{V}},
  \\
  g\big(
    J(-)
    ,\,
    J(-)
  \big)
  \;=\;
  g(-,-)

  \end{array}
\end{equation}
\vspace{-.2cm}

\noindent
uniquely {\it determine} Hermitian forms on the corresponding complex vector space $\VectorSpace{V}_{J}$,
via the formula \footnote{In symplectic geometry, the data $\big(V, J, \ImaginaryUnit g(J(-),-)\big)$ in \eqref{HermitianFormInTermsOfRealMetric} is called a {\it K{\"a}hler vector space} (e.g. \cite[p. 25]{Berndt01}) with $g$ its {\it K{\"a}hler metric} and $g\big(J(-),-\big)$ the corresponding {\it symplectic form}.}

\vspace{-.3cm}
\begin{equation}
  \label{HermitianFormInTermsOfRealMetric}
  \begin{tikzcd}
  \langle -\vert-\rangle
  \;:\,
  \VectorSpace{V}_{-J}
  \otimes_{{}_\ComplexNumbers}
  \VectorSpace{V}_{+J}
  \ar[
    r
  ]
  &[+4pt]
  \ComplexNumbers
  \,,
  \end{tikzcd}
  \hspace{.7cm}
  \langle -\vert-\rangle
    \;\;\;\defneq\;\;\;
  g(-,-) + \ImaginaryUnit g\big(J(-), -\big)
  \,,
\end{equation}
\vspace{-.4cm}

\noindent
(e.g. \cite[Lem. 1.2.15]{Huybrechts04}).
However, we highlight that under the equivalence \eqref{EquivalenceBetweenRVectorSpaceAndRealVectorBundlesOverPoint} they actually {\it become} these Hermitian forms, as follows:

\vspace{-.2cm}
\begin{equation}
  \label{InnerProductsMappedToHermitianForms}
  \begin{tikzcd}[
    row sep=0pt,
    column sep=41.5
  ]
    \big(
      \Modules{\RealNumbers},
      \otimes_{{}_{\RealNumbers}},
      \RealNumbers
    \big)
    \ar[
      rr,
      "{
        \sim
      }",
      "{
        \scalebox{.7}{
          {
            \color{darkgreen}
            \bf
            equivalence
          }
          \eqref{EquivalenceBetweenRVectorSpaceAndRealVectorBundlesOverPoint}
        }
      }"{swap}
    ]
    &&
    \big(
      \Modules{\RRealNumbers},
      \otimes_{\scalebox{.6}{$
        \RRealNumbers
      $}},
    \RRealNumbers
    \big)
    \\[12pt]
    \VectorSpace{V}
    \otimes_{{}_{\RealNumbers}}
    \VectorSpace{V}
    \ar[
      d,
      ->>,
      "{
        \mathrm{Eig}(
          J \otimes J
          ,\,
          +1
        )
      }"{description, pos=.4}
    ]
    \ar[
      rr,
      phantom,
      "{ \mapsto }"
    ]
    &&
    \big(
      \VectorSpace{V}_{-J}
      \oplus
      \VectorSpace{V}_{+J}
    \big)
    \otimes
    \big(
      \VectorSpace{V}_{-J}
      \oplus
      \VectorSpace{V}_{+J}
    \big)
    \ar[
      <->,
      bend left=60,
      shift left=7pt,
      start anchor={[xshift={-11-33pt}]},
      end anchor={[xshift={11-33pt}]},
    ]
    \ar[
      <->,
      bend left=60,
      shift left=7pt,
      start anchor={[xshift={-11+25pt}]},
      end anchor={[xshift={11+25pt}]},
    ]
    \ar[
      d,
      ->>,
      "{
        \mathrm{Eig}(
          \RealComplexStructure
          \otimes
          \RealComplexStructure
          ,\,
          +1
        )
      }"{description, pos=.4}
    ]
    \\[30pt]
    {}
    \ar[
      d,
      "{
        g
        \mathrlap{
          \raisebox{-1pt}{
          \scalebox{.7}{
            \color{darkblue}
            \bf
            \scalebox{1.3}{$\RealNumbers$}-inner product
            \color{darkgreen}
            corresponds to
          }
          }
        }
      }"{swap}
    ]
    &&
    \VectorSpace{V}_{-J}
    \otimes
    \VectorSpace{V}_{+J}
    \;\;\oplus\;\;
    \VectorSpace{V}_{+J}
    \otimes
    \VectorSpace{V}_{-J}
    \ar[
      <->,
      bend right=20,
      shift right=7pt,
      start anchor={[xshift=-13pt]},
      end anchor={[xshift=+13pt]},
    ]
    \ar[
      d,
      start anchor={[xshift=-27pt]},
      end anchor={[xshift=-5pt]},
      "{
        \mathllap{
          \raisebox{1pt}{
          \scalebox{.7}{
            \color{darkblue}
            \bf
            Hermitian form
          }
          }
        }
        \langle-\vert-\rangle
      }"{description}
    ]
    \ar[
      d,
      start anchor={[xshift=+27pt]},
      end anchor={[xshift=+5pt]},
      "{
        \langle-\vert-\rangle
        \circ
        \sigma
      }"{description}
    ]
    \\[25pt]
    \RealNumbers
    \ar[
      rr,
      phantom,
      "{ \mapsto }"
    ]
    &&
    \ComplexNumbers
    \ar[out=-45-90, in=45-90,
      looseness=3.5,
      shorten=-3pt,
      "\scalebox{.7}{${
          \hspace{5pt}
          \mathclap{
            \overline{(-)}
          }
          \hspace{5pt}
        }$}"{pos=.51, description, yshift=2pt}
    ]
  \end{tikzcd}
\end{equation}
\vspace{-.2cm}

\noindent
(This map  \eqref{InnerProductsMappedToHermitianForms} of quadratic forms
is closely related to the ``hyperbolic functor'' \cite[\S 5.2]{Bass65} of key relevance in Hermitian K-theory
\cite[p. 307]{Karoubi73}\cite[\S 3]{Bak77}\cite[\S 1.10]{Karoubi10}. For discussion relating this to KR-theory see also \cite[p. 96]{Crabb80}.)

To see that \eqref{InnerProductsMappedToHermitianForms} indeed follows from
\eqref{EquivalenceBetweenRVectorSpaceAndRealVectorBundlesOverPoint}, observe the $\RRealNumbers$-module isomorphism (cf. \cite[Lem. 1.2.5]{Huybrechts04})
\vspace{-2mm}
\begin{equation}  \label{IsomorphismFromComplexficationToHyperbolicConstruction}
  \begin{tikzcd}[
    row sep=-1pt,
   column sep=20pt
  ]
    \term{
      v + \ImaginaryUnit v'
    }
    \ar[
      rr,
      phantom, pos=.65,
      "{ \mapsto }"
    ]
    &&
    \term{
      \tfrac{1}{\sqrt{2}}
      \big(
        v - J(v')
        ,\,
        v + J(v')
      \big)
    }
    \\[+4pt]
    \ComplexNumbers
    \otimes_{{}_\RealNumbers}\!
    \VectorSpace{V}
    \ar[out=+22+180, in=-22+180,
      looseness=3,
      shift right=6pt,
      shorten <=-2pt,
      "\scalebox{1}{${
          \hspace{2pt}
          \mathclap{
            \overline{(\mbox{-})}
          }
          \hspace{4pt}
        }$}"{pos=.5, description}
    ]
    \ar[
      rr,
      <->,
      "{ \sim }"
    ]
    &&
    \VectorSpace{V}_{-J}
    \oplus
    \VectorSpace{V}_{+J}
    \ar[
      <->,
      bend left=40,
      shift left=7pt,
      start anchor={[xshift={-12-3pt}]},
      end anchor={[xshift={12-3pt}]},
    ]
    \\
    \term{
      \tfrac{1}{\sqrt{2}}
      (
        v_-
          +
        v_+
      )
      +
      \tfrac{\ImaginaryUnit}{\sqrt{2}}
      \big(
        J(v_-)
          -
        J(v_+)
      \big)
    }
    \ar[
      rr,
      phantom, pos=.28,
      "{ \mapsfrom }"
    ]
    &&
    \term{
      (v_-,\, v_+)
    }
  \end{tikzcd}
\end{equation}

\vspace{-2mm}
\noindent which serves to diagonalize
$
  J \!\otimes_{{}_\RealNumbers}\!\! \ComplexNumbers
$:
\vspace{-1mm}
\begin{equation}
  \label{TheRealComplexStructure}
  \begin{tikzcd}[
    row sep=-1pt
  ]
  \big(
    \VectorSpace{V}
      \otimes_{{}_\RealNumbers}
    \ComplexNumbers
  \big)
  \ar[
    rr,
    "{
      (
      J
        \otimes_{{}_\RealNumbers}\!
      \ComplexNumbers
      )
    }"
  ]
  \ar[
    from=d,
    "{ \sim }"{sloped}
  ]
  &&
  \big(
    \VectorSpace{V}
      \otimes_{{}_\RealNumbers}
    \ComplexNumbers
  \big)
  \ar[
    d,
    "{ \sim }"{sloped}
  ]
  \\[15pt]
   \big(
    \VectorSpace{V}_{-J}
    \oplus
    \VectorSpace{V}_{+J}
   \big)
  \ar[
    rr,
    "{
      \RealComplexStructure
    }"
  ]
  &&
   \big(
    \VectorSpace{V}_{-J}
    \oplus
    \VectorSpace{V}_{+J}
   \big)
   \\
   \term{
     (v_-
     ,\,
     v_+
     )
   }
   \ar[
     rr,
     phantom,
     "{ \mapsto }"
   ]
   &&
   \term{
     (
       -\ImaginaryUnit v
       ,\,
       + \ImaginaryUnit v
     )
     \mathrlap{\,.}
   }
  \end{tikzcd}
\end{equation}



\vspace{-1mm}
\noindent
    With \eqref{IsomorphismFromComplexficationToHyperbolicConstruction}, the isometry condition $g \circ (J \otimes J) = g$ on the left
    of \eqref{InnerProductsMappedToHermitianForms} --- which means that $g$ factors through the $(+1)$-eigenspace of
    $J \!\otimes_{{}_\RealNumbers}\! J$ --- implies by functoriality of the equivalence \eqref{EquivalenceBetweenRVectorSpaceAndRealVectorBundlesOverPoint}
    that the pairing on the right of \eqref{HermitianFormInTermsOfRealMetric} factors through the $(+1)$-eigenspace of
    $\RealComplexStructure \otimes \RealComplexStructure$ \eqref{TheRealComplexStructure}, which already makes it a Hermitian form
    on $\VectorSpace{V}_{+J}$, as shown on the right of \eqref{InnerProductsMappedToHermitianForms}.
    Explicit computation shows that this is indeed the one given by the traditional formula \eqref{HermitianFormInTermsOfRealMetric}:
    \vspace{-3mm}
    $$
    \hspace{-1mm}
  \begin{tikzcd}[
    row sep = 0pt,
    column sep=30pt
  ]
   \VectorSpace{V}_{-J}
   \otimes_{{}_\ComplexNumbers}
   \VectorSpace{V}_{+J}
   \ar[
     r,
     hook
   ]
   &
    \big(
      \VectorSpace{V}_{-J}
      \oplus
      \VectorSpace{V}_{+J}
    \big)
    \otimes_{{}_\ComplexNumbers}
    \big(
      \VectorSpace{V}_{-J}
      \oplus
      \VectorSpace{V}_{+J}
    \big)
    \ar[
      r,
      "{ \sim }",
      "{
        \scalebox{.7}{
          \color{gray}
          \scalebox{.7}{             \eqref{IsomorphismFromComplexficationToHyperbolicConstruction}
          }
        }
      }"{swap}
    ]
    &[-15pt]
    \big(
      \VectorSpace{V}
      \otimes_{{}_\RealNumbers}
      \!
      \ComplexNumbers
    \big)
    \otimes_{{}_\ComplexNumbers}
    \big(
      \VectorSpace{V}
      \otimes_{{}_\RealNumbers}
      \!
      \ComplexNumbers
    \big)
    \ar[
      r,
      "{
        (
          g
            \,\otimes_{{}_\RealNumbers}
          \ComplexNumbers
        )
        \,\otimes_{{}_\ComplexNumbers}
        (
          g
            \,\otimes_{{}_\RealNumbers}
          \ComplexNumbers
        )
      }"
    ]
    &
    \ComplexNumbers
    \\
    \scalebox{\termscale}{$
    v_-
      \otimes_{{}_\ComplexNumbers}
    v_+
    $}
     \ar[
      r,
      phantom, pos=.3,
      "{ \mapsto }"
    ]
    &
 \scalebox{\termscale}{$
 (v_-,0) \otimes_{{}_\ComplexNumbers} (0,v_+)
 $}
    \ar[
      r,
      phantom, pos=1,
      "{ \mapsto }"
    ]
    &
    \quad
\scalebox{\termscale}{$   \tfrac{1}{2}
   \big(
     v_-
     +
     \ImaginaryUnit
       J(v_-)
   \big)
   \!\otimes_{{}_\ComplexNumbers}\!
   \big(
     v_+
     -
     \ImaginaryUnit
      J(v_+)
   \big)
   $}
    \ar[
      r,
      phantom, pos=1,
      "{ \mapsto }"
    ]
   &
   \quad
\scalebox{\termscale}{$   g(v_-, v_+)
  +
  \ImaginaryUnit
  g\big(
   J(v_-)
   ,
   v_+
  \big).
  $}
    \end{tikzcd}
$$

 \noindent
{\bf Inner products with isometric complex structure internal to Real modules are Hermitian forms.}
It is instructive to restate this equivalence in the other direction: Given a Real module equipped with a symmetric
self-duality structure and an isometric complex structure (all internal to the category of Real modules!, see \cite[\S 1.4]{Monadology} for background):

\vspace{-.6cm}
\begin{equation}
  \label{SelfDualRealModulesWithIsometricComplexStructure}
  \def\arraystretch{1.2}
  \begin{array}{ccc}
    \mbox{\footnotesize
      \color{gray}
      \begin{tabular}{c}
        Real module
      \end{tabular}
    }
    &
    \begin{tikzcd}[sep=0pt]
    \ar[out=-40+180, in=40+180,
      looseness=3.3,
      shorten <=-3pt,
      shorten >=-3pt,
      "\scalebox{1}{${
          \hspace{3pt}
          \mathclap{
            C
          }
          \hspace{4pt}
        }$}"{pos=.5, description}
    ]
    \RealModule{H}
    &\in&
    \Modules{\RRealNumbers}
    \end{tikzcd}
    \\
    \mbox{\footnotesize
      \color{gray}
      \def\arraystretch{.9}
      \begin{tabular}{c}
        Symmetric
        \\
        inner product
      \end{tabular}
    }
    &
\adjustbox{raise=-10pt}{
\begin{tikzcd}
   \ComplexNumbers
    \ar[out=-41+90, in=41+90,
      looseness=3.7,
      shorten <=-2pt,
      shorten >=-3pt,
      "\scalebox{1}{${
          \hspace{3pt}
          \mathclap{
            \overline{(\mbox{-})}
          }
          \hspace{4pt}
        }$}"{pos=.5, description}
    ]
   \ar[
     r,
     "{
       \Delta
     }"
   ]
   &
    \RealModule{H}
    \otimes_{{}_\ComplexNumbers}
    \RealModule{H}
    \ar[out=-41+90, in=41+90,
      looseness=3.7,
      shorten <=-2pt,
      shorten >=-3pt,
      "\scalebox{1}{${
          \hspace{6pt}
          \mathclap{
            C \otimes C
          }
          \hspace{6pt}
        }$}"{pos=.5, description}
    ]
    \ar[
      rr,
      "{ (-,-) }"
    ]
    &&
    \ar[out=-41+90, in=41+90,
      looseness=3.7,
      shorten <=-2pt,
      shorten >=-3pt,
      "\scalebox{1}{${
          \hspace{3pt}
          \mathclap{
            \overline{(\mbox{-})}
          }
          \hspace{4pt}
        }$}"{pos=.5, description}
    ]
    \ComplexNumbers
  \end{tikzcd}
}
  &
  \def\arraystretch{1.4}
  \begin{array}{c}
    \big(
      (-) \otimes (-,-)
    \big)
    \circ
    \big(
      \Delta \otimes (-)
    \big)
    \;=\;
    \mathrm{id}_{\RealModule{H}}
    \\
    \big(
      (-,-)
        \otimes
      (-)
    \big)
    \circ
    \big(
      (-)
        \otimes
      \Delta
    \big)
    \;=\;
    \mathrm{id}_{\RealModule{H}}
    \\
    (-,-) \circ \sigma \;=\; (-,-)
  \end{array}
  \\
  \mbox{\footnotesize
    \color{gray}
    \def\arraystretch{.9}
    \begin{tabular}{c}
      Isometric
      \\
      complex structure
    \end{tabular}
  }
  &
  \begin{tikzcd}
    \RealModule{H}
    \ar[out=-41+90, in=41+90,
      looseness=3.7,
      shorten <=-2pt,
      shorten >=-3pt,
      "\scalebox{1}{${
          \hspace{3pt}
          \mathclap{
            C
          }
          \hspace{4pt}
        }$}"{pos=.5, description}
    ]
    \ar[
      rr,
      "{ \RealComplexStructure }"
    ]
    &&
    \RealModule{H}
    \ar[out=-41+90, in=41+90,
      looseness=3.7,
      shorten <=-2pt,
      shorten >=-3pt,
      "\scalebox{1}{${
          \hspace{3pt}
          \mathclap{
            C
          }
          \hspace{4pt}
        }$}"{pos=.5, description}
    ]
  \end{tikzcd}
  &
  \def\arraystretch{1.2}
  \begin{array}{c}
  \RealComplexStructure \circ \RealComplexStructure
  \,=\,
  -
  \mathrm{id}_{\RealModule{H}}
  \\
  \big(
    \RealComplexStructure(-)
    ,\,
    \RealComplexStructure(-)
  \big)
  \;=\;
  (-,-  )
  \end{array}
  \end{array}
\end{equation}
we recognize it as encoding a Hermitian form as follows.
Observing that $\RealComplexStructure$ is $\ComplexNumbers$-diagonalizable with $\mp \ImaginaryUnit$-eigenspaces swapped by the anilinear involution
(for instance by applying the previous argument \eqref{TheRealComplexStructure} to $J \,\defneq\, \RealComplexStructure^{\ZTwo}$)
\vspace{-4mm}
\begin{equation}
\label{UnderlyingSpaceOfComplexRealModule}
\begin{tikzcd}
  \HilbertSpace{H}
    \ar[out=+32+180, in=-32+180,
      looseness=4,
      shift right=4pt,
      shorten <=-0pt,
      "\scalebox{1}{${
          \hspace{2pt}
          \mathclap{
            C
          }
          \hspace{4pt}
        }$}"{pos=.5, description}
    ]
    \ar[
      rr,
      <->,
      "{ \sim }"
    ]
    &&
    \overline{\VectorSpace{H}}
    \oplus
    \VectorSpace{H}
    \mathrlap{\,,}
    \ar[
      <->,
      bend left=40,
      shift left=9pt,
      start anchor={[xshift={-8-0pt}]},
      end anchor={[xshift={8-0pt}]},
    ]
\end{tikzcd}
\end{equation}

\vspace{-4mm}
\noindent the isometry property of $\RealComplexStructure$ implies that $(-,-)$ is non-vanishing only on the $+1$ eigenspace
of $\RealComplexStructure \otimes \RealComplexStructure$, which is the mixed summands
$\begin{tikzcd}  \!\!
  \overline{\VectorSpace{H}}
  \!\otimes\!
  \VectorSpace{H}
  \;\oplus\;
  \VectorSpace{H}
  \!\otimes\!
\overline{\VectorSpace{H}}
\ar[
  <->,
  shift left=6pt,
  bend left=23,
  start anchor={[xshift={-17pt}, yshift=+1pt]},
  end anchor={[xshift={18pt}, yshift=+1pt]},
]
\end{tikzcd}$\!\!.
Here the inner product $(-,-)$ exhibits $\overline{\VectorSpace{H}}$ as the dual object (cf. \cite[\S 1]{DoldPuppe84}\cite[\S 3.1]{HeunenVicary19}) of $\VectorSpace{H}$,
inducing a $\ComplexNumbers$-linear identification of $\VectorSpace{H}$ with its anti-dual space, this defining (e.g. \cite[\S 1]{Karoubi10})
a sesqui-linear form $\langle - \vert - \rangle$:
\vspace{-2mm}
\[
  \begin{tikzcd}[row sep=-2pt]
  \VectorSpace{H}
  \ar[
    rr,
    "{ \sim }"
  ]
  &&
  \overline{\VectorSpace{H}}^\ast
  \\
  \term{
    \vert \psi \rangle
  }
  \ar[
    rr,
    phantom,
    "{ \mapsto }"
  ]
  &&
  \term{
    \langle \psi \vert
  }
  \,,
  \end{tikzcd}
\]

\vspace{-3mm}
\noindent
which is forced to be Hermitian
by symmetry and  $\ZTwo$-equivariance of $(-,-)$ (on the right, $\big\{\vert w \rangle\big\}_{w \in W}$ is any orthonormal basis):
\vspace{-2mm}
\begin{equation}  \label{CoEvaluationForHilbertSpaceAsSDObject}
  \begin{tikzcd}[row sep=10pt]
    \big(
      H
        \oplus
      H^\ast
    \big)
    \ar[
      <->,
      shift left=8pt,
      bend left=30,
      start anchor={[xshift=-8pt]},
      end anchor={[xshift=+8pt]},
    ]
    \ar[
      r,
      phantom,
      "{\mbox{$\otimes$}}"
    ]
    &[-24pt]
    \big(
      H
        \oplus
      H^\ast
    \big)
    \ar[
      <->,
      shift left=8pt,
      bend left=30,
      start anchor={[xshift=-8pt]},
      end anchor={[xshift=+8pt]},
    ]
    \ar[
      rr,
      "{
        (-,-)
      }"
    ]
    &&
    \mathbb{C}
    \ar[out=180-60, in=60,
      looseness=4.6,
      "\scalebox{1}{${
          \hspace{2pt}
          \mathclap{
            \overline{(\mbox{-})}
          }
          \hspace{4pt}
        }$}"{pos=.41, description},
        shift right=0,
        start anchor={[xshift=-0pt]},
        end anchor={[xshift=-0pt]},
    ]
    \\[-12pt]
     \term{
       \mathrlap{
        \;\;
        \langle \psi_1 \vert
      }
    }
    \ar[
      r,
      phantom,
      "{ \otimes }"
    ]
    &
    \term{
      \mathllap{
      \vert \psi_2 \rangle
      \;\;
      }
    }
    \ar[
      d,
      |->,
      shift right=22pt,
      shorten=1pt
    ]
    \ar[
      rr,
      |->,
      shorten=20pt
    ]
    &&
    \term{
      \langle \psi_1 \vert \psi_2 \rangle
    }
    \ar[
      d,
      |->,
      shorten=1pt
    ]
    \\
    \term{
    \mathrlap{
      \;\;
      \vert \psi_1 \rangle
    }
    }
    \ar[
      r,
      phantom,
      "{ \otimes }"
    ]
    &
    \term{
      \mathllap{
        \langle \psi_2 \vert
        \;\;
      }
    }
    \ar[
      r,
      |->,
      shorten >=10pt,
      shorten <=23pt
    ]
    &
    \term{
      \langle \psi_2 \vert \psi_1 \rangle
    }
    \ar[
      r,
      equals
    ]
    &
    \term{
      \overline{
        \langle \psi_1 \vert \psi_2 \rangle
      }
    }
    \\
    \term{
      \mathrlap{
        \;\;
        \vert \psi_1 \rangle
      }
    }
    \ar[
      r,
      phantom,
      "{ \otimes }"
    ]
    &
    \term{
    \mathllap{
      \vert \psi_2 \rangle
      \;\;
    }
    }
    \ar[
      d,
      |->,
      shift right=24pt,
      shorten=1pt
    ]
    \ar[
      rr,
      |->,
      shorten=25pt
    ]
    &&
    \term{ 0 }
    \ar[
      d,
      |->,
      shorten=2pt
    ]
    \\
    \term{
      \mathrlap{
        \;\;
        \langle \psi_1 \vert
      }
    }
     \ar[
      r,
      phantom,
      "{ \otimes }"
    ]
   &
    \term{
      \mathllap{
        \langle \psi_2 \vert
        \;\;
      }
    }
    \ar[
      rr,
      |->,
      shorten=25pt
    ]
    &&
   \term{ 0 }
  \end{tikzcd}
  \hspace{1cm}
  \begin{tikzcd}[row sep=10pt]
    \mathbb{C}
    \ar[out=180-60, in=60,
      looseness=4.6,
      "\scalebox{1}{${
          \hspace{2pt}
          \mathclap{
            \overline{(\mbox{-})}
          }
          \hspace{4pt}
        }$}"{pos=.41, description},
        shift right=0,
        start anchor={[xshift=-0pt]},
        end anchor={[xshift=-0pt]},
    ]
    \ar[
      rr,
      "{
        \Delta
      }"
    ]
    &&
    \big(
      H
        \oplus
      H^\ast
    \big)
    \ar[
      <->,
      shift left=8pt,
      bend left=30,
      start anchor={[xshift=-8pt]},
      end anchor={[xshift=+8pt]},
    ]
    \ar[
      r,
      phantom,
      "{\mbox{$\otimes$}}"
    ]
    &[-24pt]
    \big(
      H
        \oplus
      H^\ast
    \big)
    \ar[
      <->,
      shift left=8pt,
      bend left=30,
      start anchor={[xshift=-8pt]},
      end anchor={[xshift=+8pt]},
    ]
    \\[-8pt]
    \term{ 1 }
    \ar[
      dd,
      |->,
      shorten=12pt,
    ]
    \ar[
      rr,
      |->,
      shorten=15pt
    ]
    &&
    \term{
      \sum_w
      \;
      \vert w \rangle
    }
    \hspace{-4pt}
    \ar[
      r,
      phantom,
      "{ \otimes }"{pos=.26}
    ]
    &
    \hspace{-22pt}
    \term{
      \langle w \vert
    }
    \\[-10pt]
    &&
    \term{
       +
      \sum_w
      \;
      \langle w \vert
    }
    \;
    \ar[
      r,
      phantom,
      "{ \otimes }"{pos=.08}
    ]
    &
    \hspace{-22pt}
    \term{
      \vert w \rangle
    }
    \ar[
      d,
      |->,
      shift right=23pt,
      shorten=2pt
    ]
    \\[+10pt]
    \term{
      1
    }
    \ar[
      rr,
      |->,
      shorten=15pt
    ]
    &&
    \term{
    \sum_w
    \;
    \langle w \vert
    }
    \hspace{-4pt}
    \ar[
      r,
      phantom,
      "{ \otimes }"{pos=.26}
    ]
    &
    \hspace{-22pt}
    \term{
     \vert w \rangle
    }
    \\[-10pt]
    &&
    \term{
    +
    \sum_w
    \;
    \vert w \rangle
    }
    \;
    \ar[
      r,
      phantom,
      "{ \otimes }"{pos=.06}
    ]
    &
    \hspace{-22pt}
    \term{
     \langle w \vert
    }
  \end{tikzcd}
\end{equation}

\medskip

\noindent
{\bf The dagger emerges.}
Observe now that the {\it internally complex-linear} morphisms
$G \,:\,\HilbertSpace{H}_1 \to \HilbertSpace{H}_2$
of self-dual Real modules with internal complex structure $\RealComplexStructure$ \eqref{SelfDualRealModulesWithIsometricComplexStructure} --- hence
the Real homomorphisms $G$ satisfying $G \circ \RealComplexStructure_1 \,=\, \RealComplexStructure_2 \circ G$ --- bijectively correspond to ordinary
$\ComplexNumbers$-linear maps $g \,:\,H_1 \to H_2$ between the $(+\ImaginaryUnit)$-eigenspaces \eqref{UnderlyingSpaceOfComplexRealModule} -- because
$\ZTwo$-equivariance forces the other component to be the Hermitian adjoint operator $g^\dagger$, defined by
$\langle - \vert g^\dagger - \rangle \,=\, \langle g - \vert - \rangle$ (e.g. \cite[(A.15)]{Landsman17}):
\vspace{-2mm}
  \begin{equation}   \label{ComplexLinearMapsAsIBetaComplexLinearRealHomomorphisms}
    \adjustbox{raise=4pt}{
      \begin{tikzcd}[
        row sep=-2pt,
        column sep=15pt
      ]
        \mathllap{
          g
          \;\isa\;\;
        }
        H_1
        \ar[
          rr,
          "{
            \scalebox{.7}{
              \rm
              \scalebox{1.2}{$\mathbb{C}$}-linear map
            }
          }"
        ]
        &&
        H_2
        \\
        \term{
          \vert \psi \rangle
        }
        &\mapsto&
        \term{
          g \vert \psi \rangle
        }
      \end{tikzcd}
      }
    \hspace{1.6cm}
    \leftrightarrow
    \hspace{1.6cm}
    \begin{tikzcd}[
      column sep=40pt,
      row sep=5pt
    ]
      \mathllap{
        G
        \;\;\isa\;\;
      }
      \HilbertSpace{H}_1
      \ar[
        rr,
        "{
          \scalebox{.8}{
            \rm \footnotesize
            \def\arraystretch{.9}
            \begin{tabular}{c}
            \scalebox{1.3}{$\RealComplexStructure$}-complex-linear
            \\
            Real-module homom.
            \end{tabular}
          }
        }"
      ]
      &\phantom{----}&
      \HilbertSpace{H}_2
      \\[-4pt]
      \term{
        \vert \psi \rangle
      }
      \ar[
        dd,
        |->,
        shorten=2pt
      ]
      \ar[
        rr,
        |->,
        shorten=25pt
      ]
      &&
      \term{
        g\vert \psi \rangle
      }
      \ar[
        dd,
        |->,
        shorten=2pt
      ]
      \\
      \\
      \term{
        \langle \psi \vert
      }
      \ar[
        r,
        |->,
        shorten=2pt
      ]
      &
      \term{
        \langle \psi \vert
        \,
       {\color{purple}g^\dagger}
      }
      \ar[
        r,
        equals
      ]
      &
      \term{
        \langle g \psi \vert
      }
      \mathrlap{\,.}
    \end{tikzcd}
  \end{equation}

  \vspace{-2mm}
\noindent
In particular, the {\it internally isometric} internally-complex-linear Real module homomorphisms correspond
bijectively to the complex isometries on the underlying Hermitian spaces:
\vspace{-2mm}
\begin{equation}
  \label{InternalIsometryProperty}
  \begin{tikzcd}[row sep=-3pt, column sep=2pt]
    \HilbertSpace{H}_1
    \otimes
    \HilbertSpace{H}_1
    \ar[
      rrrr,
      "{
        G \,\otimes\, G
      }"
    ]
      \ar[
        dddd,
        "{ (-,-)_1 }"
        {swap}
      ]
    &&&&
    \HilbertSpace{H}_2
    \otimes
    \HilbertSpace{H}_2
      \ar[
        dddd,
        "{ (-,-)_2 }"
      ]
    \\
    &
    \term{
      \langle
       \phi
      \vert
      \otimes
      \vert \psi
      \rangle
    }
      \ar[
        dd,
        phantom,
        "{ \longmapsto }"{sloped}
      ]
    &\longmapsto&
    \term{
      \langle
       \phi
      \vert g^\dagger
      \otimes
      g \vert \psi
      \rangle
    }
    \\
    &
    \rotatebox[origin=c]{-90}{$\longmapsto$}
    &&
    \rotatebox[origin=c]{-90}{$\longmapsto$}
    \\
    &
   \term{ \langle \phi \vert \psi \rangle }
    \ar[
      rr,
      equals
    ]
    &&
    \term{
      \langle \phi
      \vert
      g^\dagger g
      \vert
      \psi \rangle
    }
    \\
    \RRealNumbers
    \ar[
      rrrr,
      equals
    ]
    &&&&
    \RRealNumbers
  \end{tikzcd}
\end{equation}

\vspace{-2mm}
\noindent hence to unitary operators iff they are also invertible; and the dagger-functor on the category of
Hermitian spaces is now just the duality-involution (cf. eg. \cite[Def. 3.11]{HeunenVicary19}) on this category of self-dual Real modules, as shown here:
\vspace{-2mm}
\begin{equation}
  \label{HermitianAdjointFromSelfDuality}
  \begin{tikzcd}[
    row sep=-5pt,
    column sep=10pt
  ]
    &
    &
    \HilbertSpace{H}_1
    \ar[
      rr,
      equals
    ]
    &&
    \HilbertSpace{H}_1
    \ar[
      rr,
      equals
    ]
    &&
    \HilbertSpace{H}_1
    \\
    &
    \RRealNumbers
    \ar[
      r,
      "{
        \Delta
      }"
    ]
    & \otimes
    &&
    \otimes
    \\
    &
    &
    \HilbertSpace{H}_1
    \ar[
      rr,
      "{ G }"
    ]
    &\phantom{---}&
    \HilbertSpace{H}_2
    \\
    & &
    \otimes
    &&
    \otimes
    \ar[
      r,
      "{
        (-,-)
      }"
    ]
    &
    \RRealNumbers
    \\
    \HilbertSpace{H}_2
    \ar[
      rr,
      equals
    ]
    &&
    \HilbertSpace{H}_2
    \ar[
      rr,
      equals
    ]
    &&
    \HilbertSpace{H}_2
    \\[+5pt]
    \term{
      \vert \psi \rangle
    }
    \ar[
      rr,
      phantom,
      "{ \mapsto }"
    ]
    &
    &
    \def\arraystretch{.9}
    \begin{array}{r}
    \term{
    \underset{
      w
    }{\sum}
    \,
    \vert w \rangle
    \otimes
    \langle w \vert
    \otimes
    \vert \psi \rangle
    }
    \\
    \term{
    + \,
    \underset{
      w
    }{\sum}
    \,
    \langle w \vert
    \otimes
    \vert w \rangle
    \otimes
    \vert \psi \rangle
    }
    \end{array}
    \ar[
      rr,
      phantom,
      "{ \mapsto }"
    ]
    &&
    \def\arraystretch{.9}
    \begin{array}{r}
    \term{
    \underset{
      w
    }{\sum}
    \,
    \vert w \rangle
    \otimes
    \langle w \vert \, g^\dagger
    \otimes
    \vert \psi \rangle
    }
    \\
    \term{
    +\,
    \underset{
      w
    }{\sum}
    \,
    \langle w \vert
    \otimes
    \; g \, \vert w \rangle
    \otimes
    \vert \psi \rangle
    }
    \end{array}
    \ar[
      rr,
      phantom,
      "{ \mapsto }"
    ]
    &&
    \term{
    \underset{
      {
        \color{purple}
        g^\dagger
      }
      \,
      \vert \psi \rangle
    }{
    \underbrace{
      \underset{w}{\textstyle{\sum}}
      \vert w \rangle
      \langle w \vert G^\dagger \vert \psi \rangle
    }}
    }
  \end{tikzcd}
\end{equation}

\vspace{-2mm}
\noindent
{\bf Hermitian operators as Real matrices.} Similarly, we find that ``density matrices'', hence  Hermitian operators on (finite-dimensional, in our case)
Hilbert spaces (see \cite[\S 1.2]{Monadology} for background) are just the complex {\it symmetric} matrices internal to Real modules.
The Real space $\mathrm{CSMat}(\HilbertSpace{H})$ of internally-complex (C) and symmetric (S) matrices on an internally complex self-dual
Real space $\HilbertSpace{H}$ is the following iterated equalizer:
\vspace{-2mm}
$$
  \begin{tikzcd}[column sep=large]
    \mathrm{CSMat}(\HilbertSpace{H})
    \ar[r]
    \ar[d]
    \ar[
      dr,
      phantom,
      "{
        \scalebox{.7}{(pb)}
      }"
    ]
    &
    \mathrm{CMat}(\HilbertSpace{H})
    \ar[
      d,
      hook,
      "{ \mathrm{eq} }"
    ]
    \ar[
      rr,
      dashed,
      "{
         \braiding
           { \otimes }
           {
             \mathrm{CMat}(\HilbertSpace{H})
           }
      }"
    ]
    &&
    \mathrm{CMat}(\HilbertSpace{H})
    \ar[
      d,
      hook,
      "{ \mathrm{eq} }"
    ]
    \\
    \mathrm{SMat}(\HilbertSpace{H})
    \ar[
      r,
      hook,
      "{ \mathrm{eq} }"
    ]
    \ar[
      dd,
      dashed,
      "{
        \RealComplexStructure
        \,\otimes\,
        \RealComplexStructure_{
          \vert
          \mathrm{SMat}(\HilbertSpace{H})
        }
      }"{description}
    ]
    &
    \HilbertSpace{H}
    \otimes
    \HilbertSpace{H}
    \ar[
      rr,
      shift left=6pt,
      "{
        \braiding
          { \otimes }
          {
            \HilbertSpace{H}
            \otimes
            \HilbertSpace{H}
          }
      }"{description}
    ]
    \ar[
      rr,
      shift right=6pt,
      "{
        \mathrm{id}
      }"{description}
    ]
    \ar[
      dd,
      shift left=7pt,
      "{
        \RealComplexStructure
        \,\otimes\,
        \RealComplexStructure
      }"{description}
    ]
    \ar[
      dd,
      shift right=7pt,
      "{ \mathrm{id} }"{description}
    ]
    &&
    \HilbertSpace{H}
    \otimes
    \HilbertSpace{H}
    \ar[
      dd,
      shift left=7pt,
      "{
        \RealComplexStructure
        \,\otimes\,
        \RealComplexStructure
      }"{description}
    ]
    \ar[
      dd,
      shift right=7pt,
      "{ \mathrm{id} }"{description}
    ]
    \\
    \\
    \mathrm{SMat}(\HilbertSpace{H})
    \ar[
      r,
      hook,
      "{ \mathrm{eq} }"
    ]
    &
    \HilbertSpace{H}
    \otimes
    \HilbertSpace{H}
    \ar[
      rr,
      shift left=6pt,
      "{
        \braiding
          { \otimes }
          {
            \HilbertSpace{H}
            \otimes
            \HilbertSpace{H}
          }
      }"{description}
    ]
    \ar[
      rr,
      shift right=6pt,
      "{
        \mathrm{id}
      }"{description}
    ]
    &&
    \HilbertSpace{H}
    \otimes
    \HilbertSpace{H}
  \end{tikzcd}
$$
Unwinding the definitions, one has
\[
  \begin{tikzcd}[
    row sep=0pt
  ]
    \mathllap{
      \mathrm{CMat}(\HilbertSpace{H})
      \;\simeq\;\;
    }
    \big(
      H
        \!\otimes_{{}_\ComplexNumbers}\!
      H^\ast
      \,\oplus\,
      H^\ast
        \!\otimes_{{}_\ComplexNumbers}\!
      H
    \big)
    \ar[
      <->,
      bend left=25pt,
      shift left=9pt,
      start anchor={[xshift=-22pt]},
      end anchor={[xshift=+20pt]},
    ]
    \ar[
      rr,
      "{
        \braiding
          { \otimes }
          {
            \mathrm{CMat}(\HilbertSpace{H})
          }
      }"
    ]
    &&
    \big(
      H
        \!\otimes_{{}_\ComplexNumbers}\!
      H^\ast
      \,\oplus\,
      H^\ast
        \!\otimes_{{}_\ComplexNumbers}\!
      H
    \big)
    \ar[
      <->,
      bend left=25pt,
      shift left=9pt,
      start anchor={[xshift=-22pt]},
      end anchor={[xshift=+20pt]},
    ]
    \\
    \term{
      \vert \psi \rangle
      \otimes
      \langle \phi \vert
      {
      \;\;
      \color{gray}\leftrightarrow\;
      \langle \psi \vert
      \otimes
      \vert \phi \rangle
      }
    }
    \ar[
      rr,
      phantom,
      "{ \longmapsto }"
    ]
    &&
    \term{
      {
      \color{gray}\vert \phi \rangle
      \otimes
      \langle \psi \vert
      \;\leftrightarrow\;
      }
      \langle \phi \vert
      \otimes
      \vert \psi \rangle
    }
    \mathrlap{\,,}
  \end{tikzcd}
\]
which means that the $\ZTwo$-fixed locus of $\mathrm{CMat}(\HilbertSpace{H})$ is identified with the $\RealNumbers$-vector space underlying
$H \!\otimes_{{}_\ComplexNumbers}\! H^\ast$, under which identification the braiding acts as the Hermitian adjoint operation
$(-)^\dagger \,:\, \vert \psi \rangle \langle \phi \vert \,\mapsto\, \vert \phi \rangle \langle \psi \vert$:

\newpage
$\,$

\vspace{-7mm}
\[
  \begin{tikzcd}[
    row sep=-2pt, column sep=small
  ]
    \mathrm{CMat}(\HilbertSpace{H})^{\ZTwo}
    \ar[
      dddd,
      hook
    ]
    &[+10pt]
    H
      \otimes_{{}_\ComplexNumbers}
    H^\ast
    \ar[
      rrrr
    ]
    \ar[
      dddd,
      hook
    ]
    &[-30pt]
    &&
    &[-30pt]
    H
      \otimes_{{}_\ComplexNumbers}
    H^\ast
    \ar[
      dddd,
      hook
    ]
    \\
    &
    &
    \term{
      \vert \psi \rangle
      \langle \phi \vert
    }
    &\longmapsto&
    \term{
      \vert \phi \rangle
      \langle \psi \vert
    }
    \\
    &
    &
    \rotatebox[origin=c]{-90}{$\longmapsto$}
    &&
    \rotatebox[origin=c]{-90}{$\longmapsto$}
    \\
    &
    &
    \term{
      \vert \psi \rangle
      \otimes
      \langle \phi \vert
      \;+\;
      \langle \psi \vert
      \otimes
      \vert \phi \rangle
    }
    &\longmapsto&
    \term{
      \vert \phi \rangle
      \otimes
      \langle \psi \vert
      \;+\;
      \langle \phi \vert
      \otimes
      \vert \psi \rangle
    }
    \\
    \mathrm{CMat}(\HilbertSpace{H})
    &
    \big(
      H
        \!\otimes_{{}_\ComplexNumbers}\!
      H^\ast
      \,\oplus\,
      H^\ast
        \!\otimes_{{}_\ComplexNumbers}\!
      H
    \big)
    \ar[
      rrrr,
      "{
        \braiding
          { \otimes }
          {
            \mathrm{CMat}(\HilbertSpace{H})
          }
      }"{swap}
    ]
    &&&&
    \big(
      H
        \!\otimes_{{}_\ComplexNumbers}\!
      H^\ast
      \,\oplus\,
      H^\ast
        \!\otimes_{{}_\ComplexNumbers}\!
      H
    \big)
  \end{tikzcd}
\]

\vspace{-1.5mm}
\noindent This reveals $\mathrm{CSMat}(\HilbertSpace{H})$ as the Real space corresponding under
\eqref{EquivalenceBetweenRVectorSpaceAndRealVectorBundlesOverPoint} to the $\RealNumbers$-vector space of Hermitian operators:
\vspace{-4mm}
\[
  \mathrm{CSMat}(\HilbertSpace{H})
  \;\simeq\;
  \mathrm{Herm}\big(
    H
    ,\,
    \langle\mbox{-}\vert\mbox{-}\rangle
  \big)
  \otimes_{{}_\RealNumbers}
  \hspace{-6pt}
  \begin{tikzcd}
    \ComplexNumbers
    \ar[out=-40, in=40,
      looseness=3.4,
      shorten <=-3pt,
      shift left=3pt,
      "\scalebox{1}{${
          \hspace{2pt}
          \mathclap{
            \overline{(\mbox{-})}
          }
          \hspace{4pt}
        }$}"{pos=.5, description}
    ]
  \end{tikzcd}
\]

\vspace{-5mm}
\noindent The Real incarnation of a unitary quantum channel induced by a unitary
operator \eqref{ComplexLinearMapsAsIBetaComplexLinearRealHomomorphisms} is hence simply:
\vspace{-2mm}
$$
  \begin{tikzcd}[row sep=0pt,
    column sep=large
  ]
  \term{
    \vert \psi \rangle
    \langle \phi \vert
   }
   &\longmapsto&
   \term{
     g\,
     \vert \psi \rangle
     \langle \phi \vert
     \,
     g^\dagger
   }
   \\
   \HilbertSpace{H}
     \otimes
   \HilbertSpace{H}
    \ar[
      rr,
      "{
        G \,\otimes\, G
      }"
    ]
    &&
    \HilbertSpace{H}
      \otimes
    \HilbertSpace{H}
    \\[15pt]
    \mathrm{CSMat}(\HilbertSpace{H})
    \ar[
      rr
    ]
    \ar[
      u,
      hook
    ]
    &&
    \mathrm{CSMat}(\HilbertSpace{H})
    \ar[
      u,
      hook
    ]
    \\[-3pt]
    \term{
      \rho
    }
      &\longmapsto&
    \term{
      g \cdot \rho \cdot g^\dagger
    }
  \end{tikzcd}
$$

\noindent
{\bf In conclusion so far}, the above shows that the Hermitian structures in the foundations of quantum information theory are naturally expressed by
isometrically complex self-dual objects \eqref{SelfDualRealModulesWithIsometricComplexStructure} {\it internal} to the category of Real modules.
Since the latter is monoidally equivalent \eqref{EquivalenceBetweenRVectorSpaceAndRealVectorBundlesOverPoint} to that of $\RealNumbers$-vector spaces,
the same is already true there, where the above constructions amount to strictly referring to Hermitian forms only through their real part
\eqref{HermitianFormInTermsOfRealMetric} regarded as a symmetric form on the underlying $\RealNumbers$-vector space \eqref{RealPartOfHermitianForm}
-- which is an elementarily equivalent but unusual perspective.
It is only when that same construction is transported back through the equivalence with Real modules \eqref{EquivalenceBetweenRVectorSpaceAndRealVectorBundlesOverPoint}
that all the familiar $\ComplexNumbers$-component formulas of quantum theory manifest themselves, such as for Hermitian adjoints \eqref{HermitianAdjointFromSelfDuality}.

However, abstractly, such as for formal languages, the component expressions are invisible/irrelevant anyway -- they are part of the ``semantics''
(the model) but not of the ``syntax'' (the theory). One may understand these observations as one way of seeing ``why'' the (spectral) theory
of Hermitian forms on $\ComplexNumbers$-vector famously parallels that of symmetric inner products on $\RealNumbers$-vector spaces:
They secretly {\it are} symmetric inner products when seen internally to Real modules.

\smallskip

In amplification of this point, we now  indicate how one may formalize (finite-dimensional) Hilbert space structure internal to a formal language
like {\LHoTT}, without introducing further inference rules for anti-linear maps.

\medskip

\noindent
{\bf Formalization in linearly homotopy-typed languages.}
Looking back, one may notice that the only property of the tensor unit $\TensorUnit \,\defneq\, \RealNumbers$ in the symmetric closed monoidal
category $(\mathcal{C}, \otimes, \TensorUnit, \sigma) \,\defneq\,(\Modules{\RealNumbers}, \otimes_{{}_\RealNumbers}, \RealNumbers, \sigma_{{}_\RealNumbers})$
that we used above --- namely for saying ``internal complex structure'' in \eqref{RealPartOfHermitianForm} and
\eqref{SelfDualRealModulesWithIsometricComplexStructure} --- is that it contains a negative unit scalar, in the sense of a non-trivial involutive endomorphism:
\begin{equation}
  \label{NegativeUnitScalar}
  -\mathrm{id}
  \,:\,
  \TensorUnit
  \to
  \TensorUnit
  ,\;\;
  \hspace{.5cm}
  \big(-\mathrm{id} = \mathrm{id}\big)
  \,\to\,
  \varnothing
  ,\hspace{.4cm}
  -\mathrm{id}
  \circ
  -\mathrm{id}
  \,=\,
  \mathrm{id}
  \,.
\end{equation}
Trivial as this may seem, it is not generally the case for tensor units of monoidal categories and it needs an argument that a term like
\eqref{NegativeUnitScalar} is constructible in a given formal language for such categories. Interestingly, that this is the case
for {\LHoTT} relies on its homotopy-theoretic nature. In indicating now how this works, we will here not go into the formal syntax
of {\LHoTT}, but discuss the relevant universal constructions in its expected categorical sematics which interpret this term:

\smallskip

Namely, in refinement of how plain {\HoTT} has categorical semantics in general $\infty$-toposes
(see \cite[p. 45]{TQP} for pointers)\footnote{We will notationally suppress the ``$\infty$-''-prefix in the following.
All constructions are understood to be homotopy-theoretic.
Similarly, all diagrams are filled by 2-morphisms, even when these are not explicitly indicated.} and expresses the universal constructions
that are generically available in all of them,
so {\LHoTT} is meant to have categorical semantics in ``tangent $\infty$-toposes'' of parameterized $R$-module spectra
(for $R$ an $E_\infty$-ring, for pointers see \cite[\S 1.5]{Monadology}) and to express the universal constructions generically available in all of these.

Since such tangent $\infty$-topos is meant to be (categorical semantics for) the type universe of {\LHoTT} we suggestively denote it by ``$\Types$'',
as in \cite{Monadology}, with the reflective sub-$\infty$-categories of purely classical types (homotopy types, $\infty$-groupoids)
and of purely linear (quantum) types (module spectra) denoted

\vspace{-3mm}
\begin{equation}
  \label{TheToposOfTypes}
  \begin{tikzcd}[
    row sep=0pt, column sep=large
  ]
    \ClassicalTypes
  \quad   \ar[
      r,
      hook,
      shift right=5pt
    ]
    \ar[
      from=r,
      ->>,
      shift right=5pt,
      "{ \classically }"{swap}
    ]
    \ar[
      r,
      phantom,
      "{
        \scalebox{.7}{$\bot$}
      }"
    ]
    &
  \quad   \Types \quad
    &
  \quad   \QuantumTypes
    \ar[
      l,
      hook',
      shift left=5pt
    ]
    \ar[
      from=l,
      ->>,
      shift left=5pt,
      "{ \quantumly }"
    ]
    \ar[
      l,
      phantom,
      "{
        \scalebox{.7}{$\bot$}
      }"
    ]
    \\
 \scalebox{\termscale}{$    \left\{\,
        \ABundleType
      { 0 }
      { X }
       \, \right\}
    $}
    &
  \scalebox{\termscale}{$   \left\{\,
    \ABundleType
      { \VectorSpace{V}_\bullet }
      { X }
   \, \right\}
    $}
    &
  \scalebox{\termscale}{$   \left\{\,
    \ABundleType
      { \VectorSpace{V} }
      { \ast }
       \, \right\}
        $}
    \mathrlap{\,.}
  \end{tikzcd}
\end{equation}
\vspace{-.4cm}

\medskip

\noindent
{\bf The doubly symmetric closed monoidal structure.}
Here
\begin{itemize}[leftmargin=.5cm]
\item
$\ClassicalTypes$ is Cartesian monoidal with Cartesian product denoted $\times$ and tensor unit denoted $\ast$,
\item
$\QuantumTypes$ is symmetric monoidal with tensor product $\otimes$ and tensor unit $\TensorUnit$, distributing over
a  Cartesian biproduct (direct sum) $\oplus$ with tensor unit the zero object,
\item $\Types$ is doubly monoidal with the corresponding ``external'' Cartesian and tensor products (cf. \cite{EoS},
here to be denoted by the same symbols $\times$ and $\otimes$, respectively),
\end{itemize}
such that all functors in \eqref{TheToposOfTypes} are strong monoidal.

Moreover, $\QuantumTypes$ is stable (in the sense of \cite[\S 1]{Lurie17}, namely under suspension), meaning that
for $\VectorSpace{V} \,\in\, \QuantumTypes$ the canonical comparison map to the looping $\Omega(-)$ of
its suspension $\Sigma(-)$ is an equivalence (cf. \cite[p. 24]{Lurie17}):

\vspace{-3mm}
\begin{equation}
  \begin{tikzcd}[row sep=7pt, column sep=20pt]
    &
    0
    \ar[
      dr,
      shorten=-2pt
    ]
    \ar[
      dd,
      Rightarrow,
      shorten=4pt,
      "{
        \scalebox{.7}{(po)}
      }"{description, pos=.47}
    ]
    \\
    V
    \ar[
      ur,
      shorten=-2pt
    ]
    \ar[
      dr,
      shorten=-2pt
    ]
    &&
    \Sigma V
    \mathrlap{\,,}
    \\
    &
    0
    \ar[
      ur,
      shorten=-2pt
    ]
  \end{tikzcd}
  \hspace{15pt}
  \begin{tikzcd}[sep=7pt, column sep=20pt]
    &
    0
    \ar[
      dr,
      shorten=-2pt
    ]
    \ar[
      dd,
      Rightarrow,
      shorten=4pt,
      "{
        \scalebox{.7}{(pb)}
      }"{description, pos=.47}
    ]
    \\
    \Omega V
    \ar[
      ur,
      shorten=-2pt
    ]
    \ar[
      dr,
      shorten=-2pt
    ]
    &&
    V
    \mathrlap{\,,}
    \\
    &
    0
    \ar[
      ur,
      shorten=-2pt
    ]
  \end{tikzcd}
  \hspace{15pt}
  \begin{tikzcd}[
    column sep=20pt,
    row sep=4pt
  ]
    &[8pt]
    &[-8pt]
    0
    \ar[ddrr]
    \\
    \\
    V
    \ar[
      uurr,
      bend left=10pt
    ]
    \ar[
      ddrr,
      bend right=10pt
    ]
    \ar[
      r,
      dashed,
      "{ \sim }"
    ]
    &
    \Omega \Sigma V
    \ar[
      uur,
      end anchor={[xshift=3.5pt]}
    ]
    \ar[
      ddr,
      end anchor={[xshift=3.5pt]}
    ]
    &&&
    \Sigma V \, .
    \\
    \\
    &&
    0
    \ar[uurr]
  \end{tikzcd}
\end{equation}

Of interest for our application to quantum information theory is the case where the ground $E_\infty$-ring is the
real Eilenberg-MacLane spectrum $R \,\defneq\, H\RealNumbers$, in which case $\Types$ is equivalently the flat
$\infty$-vector bundles or $\infty$-local systems, see \cite{EoS}, hence $\QuantumTypes$
is equivalently given by $\RealNumbers$-chain complexes; see \eqref{HeartOfHRSpectra} below.

\medskip

\noindent
{\bf The negative unit scalar in {\LHoTT}.}
If we denote the homotopy exhibiting the suspension type (cf. \cite[(117)]{TQP}) of the tensor unit $\TensorUnit$ by $s$,
then its inverse $s^{-1}$ induces an endomorphism of $\Sigma \TensorUnit$ (by the universal property of the pushout)
whose looping is the scalar which we may denote by ``$-\mathrm{id}$'':

\vspace{-6mm}
\begin{equation}
  \label{TheNegativeUnitScalar}
  \hspace{-7mm}
  \begin{tikzcd}[
    column sep=25pt,
    row sep=10pt
  ]
    &
    0
    \ar[
      dr,
      shorten=-2pt
    ]
    \ar[
      dd,
      Rightarrow,
      shorten=4pt,
      "{ s }",
      "{
        \scalebox{.65}{(po)}
      }"{swap}
    ]
    \\
    \TensorUnit
    \ar[
      ur,
      shorten=-2pt
    ]
    \ar[
      dr,
      shorten=-2pt
    ]
    &&
    \Sigma \TensorUnit
    \\
    &
    0
    \ar[
      ur,
      shorten=-2pt
    ]
  \end{tikzcd}
  \hspace{1cm}
  \begin{tikzcd}[
    column sep=30pt,
    row sep=6pt
  ]
    &[-10pt]
    &
    0
    \ar[
      ddr,
      "{\ }"{swap, pos=.1, name=s1}
    ]
    \ar[
      ddrr,
      bend left=10pt,
      "{\ }"{swap, pos=.2, name=s2}
    ]
    &&[+10pt]
    \\
    \\
    \TensorUnit
    \ar[uurr]
    \ar[
      ddrr,
      "{\ }"{name=t1, pos=.8}
    ]
    &&&
    \Sigma \TensorUnit
    \ar[
      r,
      dashed,
      "{ \Sigma(-\mathrm{id}) }"{pos=.35},
      "{ \sim }"{swap, pos=.35}
    ]
    &
    \Sigma \TensorUnit
    \mathrlap{\,,}
    \\
    \\
    &&
    0
    \ar[
      uurr,
      bend right=10pt,
      "{\ }"{pos=.2, name=t2}
    ]
    \ar[
      uur
    ]
    \ar[
      from=s1,
      to=t1,
      Rightarrow,
      "{ s }"{description}
    ]
    \ar[
      from=s2,
      to=t2,
      Rightarrow,
      crossing over,
      "{ s^{-1} }"{description}
    ]
  \end{tikzcd}
  \hspace{1cm}
  -\mathrm{id}
  \,=\,
  \Omega \Sigma(-\mathrm{id})
  \;:\;
  \TensorUnit
  \to
  \TensorUnit
  \mathrlap{\,.}
\end{equation}

\vspace{-2mm}
\noindent The desired characteristic properties \eqref{NegativeUnitScalar}
of this $-\mathrm{id}$ \eqref{TheNegativeUnitScalar} follow immediately from similar uses of the universal property of the defining pushouts.
In the language of higher algebra, this is the non-trivial element of degree=0 in $\mathrm{GL}(1, \mathbb{S})$,  the ``group of units
of the sphere spectrum'' (cf. \cite[\S 1.2]{ABGHR14}\cite[(3.18)]{SSS23Character}):
$\pi_0\big(\mathrm{GL}(1,\mathbb{S})\big) = \big(\pi_0\mathbb{S}\big)^\times \,=\, \Integers^\times \,=\, \ZTwo$.

\medskip

\noindent
{\bf The heart of {\LHoTT}.} With the negative unit scalar \eqref{TheNegativeUnitScalar} in hand, it is immediate to define data
structures (see \cite[p. 53]{TQP} for background) in $\QuantumTypes$ of symmetric inner product spaces with isometric complex
structure according to \eqref{SelfDualRealModulesWithIsometricComplexStructure}. The only further subtlety to take care of
is that the definition \eqref{SelfDualRealModulesWithIsometricComplexStructure} comes out as intended (only) on linear base
types which are in the ``heart'' of $\QuantumTypes$. For completeness, we briefly explain this.
In plain {\HoTT} the $n$-truncation modality (for $n \in \mathbb{N}$, see \cite[p. 50]{TQP} for pointers) sends a homotopy type
$X \,:\, \Types$ to a type $[X]_n \,:\, \Types$ which left-universally approximates $X$ while containing no non-trivial $(n+1)$-fold
homotopies. In particular, the ordinary definitions of mathematical structures, such as of groups (cf. \cite[(140)]{TQP}), come out
as intended (only) on 0-truncated base types $X = [X]_0$. Instead, for higher base types these definitions need to be refined
to ``higher structures''. While higher structures are profoundly interesting --- and while we have here at our fingertips a
definition of higher ``$(\infty,1)$-Hilbert spaces'' (of finite type) which is worth exploring further --- for the scope
of this note we want to restrict attention to ordinary quantum state spaces.

\smallskip

To that end, the further subtlety in {\LHoTT} is that in a stable $\infty$-category like the $\QuantumTypes$, the plain notion
of truncation becomes meaningless: For all $n \in \mathbb{N}$, the {\it only} object of $\QuantumTypes$ which is $n$-truncated
is the 0-object (\cite[Warning 1.2.1.9]{Lurie17}). Instead, the proper notion that replaces $n$-truncation in stable
$\infty$-categories are ``t-structures'' (\cite[\S 1.2.1]{Lurie17}), and the stable (i.e., linear) analog of the
0-truncated sector is the ``heart'' of the t-structure (\cite[Def. 1.2.1.11]{Lurie17}).

\smallskip
Concretely, the heart of {\LHoTT} may be characterized as follows. Consider the function which sends a quantum type to
\vspace{-2mm}
\[
  \Omega^\infty
  \;\defneq\;
  \classically\big(
    \TensorUnit
    \to
    (-)
  \big)
  \;:\;
  \QuantumTypes
  \longrightarrow
  \ClassicalTypes
  \,,
\]
\vspace{-2mm}

\noindent
the classicalization \eqref{TheToposOfTypes} of its internal hom out of
the tensor unit (cf. \cite[Def. 2.1.26]{Riley22}\cite[Prop. 2.7]{Monadology}). Then a linear type belongs to the heart iff
\begin{itemize}
\item[{\bf (i)}]
under classicization $\classically$ \eqref{TheToposOfTypes} it is identified with its 0-truncation,
\item[{\bf (ii)}]
the 0-truncations of the classicization of all its suspensions are contractible:
(cf. \cite[Prop. 1.4.3.4]{Lurie17}): \footnote{
In terms of module spectra this may be readily understood from the fact that the stable homotopy groups
of the suspensions of a spectrum relate to those of its underlying homotopy type by
$
  \pi_k\big(
    \Omega^\infty
    \Sigma^n
    \VectorSpace{V}
  \big)
  \,
  =
  \,
  \pi_{n-k}\big(
    \VectorSpace{V}
  \big)
  \,.
$
In view of this, the heart condition \eqref{ConditionForTheHeart} says equivalently that the $\Integers$-graded stable homotopy
groups $\pi_\bullet(\VectorSpace{V})$ are concentrated in degree 0.
}
\end{itemize}

\vspace{-2mm}
\begin{equation}
  \label{ConditionForTheHeart}
  \QuantumTypes^\heart
  \;\;
  \defneq
  \;\;
  \Big(\,
  V \,:\, \QuantumTypes
  ,\;\;\;
  \Omega^\infty \VectorSpace{V} \,=\, [\Omega^\infty \VectorSpace{V}]_0
  ,\;\;\;
  \underset{n \in \mathbb{N}}{\forall}\; [\Omega^\infty \Sigma^n \VectorSpace{V}]_0 \,=\, \ast
 \, \Big)\, .
\end{equation}

\newpage

In particular, the heart of $H\RealNumbers$-module-spectra is the ordinary category of $\RealNumbers$-vector spaces:
\vspace{-2mm}
\begin{equation}
  \label{HeartOfHRSpectra}
   \big(
     \Modules{H\RealNumbers}
     ,\,
     \otimes_{{}_{H\RealNumbers}}
     ,\,
     H\RealNumbers
     ,\,
     \sigma_{{}_{H\RealNumbers}}
  \big)^\heart
  \;\simeq\;
   \big(
     \Modules{\RealNumbers}
     ,\,
     \otimes_{{}_{\RealNumbers}}
     ,\,
     \RealNumbers
     ,\,
     \sigma_{{}_{\RealNumbers}}
  \big)
  \,.
\end{equation}

\noindent
{\bf In conclusion}, for the purpose of using {\LHoTT} as a universal quantum programming and certification language (as laid out in \cite{Monadology}, in certifiable refinement of the Proto-{\Quipper} language),
one may code (finite-dimensional) Hilbert spaces in {\LHoTT} as heart-types \eqref{ConditionForTheHeart}
among purely linear types \eqref{TheToposOfTypes}
equipped with symmetric self-duality and isometric complex structure as in \eqref{SelfDualRealModulesWithIsometricComplexStructure},
using the negative unit scalar type \eqref{TheNegativeUnitScalar}.
When interpreted into $H\RealNumbers$-module spectra (as in \cite{EoS}), this gives, up to equivalence \eqref{InnerProductsMappedToHermitianForms},
the ordinary symmetric closed monoidal category of (finite-dimensional) $\ComplexNumbers$-Hilbert spaces with isometries
between them, whose dagger-structure is induced by plain dualization \eqref{HermitianAdjointFromSelfDuality}.

\smallskip

\noindent
{\bf Outlook.} The restriction to the heart of {\LHoTT} in \eqref{ConditionForTheHeart} serves the purpose (only) of showing that traditional quantum information theory does embed faithfully into {\LHoTT}. If we just drop this constraint then the language construct \eqref{SelfDualRealModulesWithIsometricComplexStructure} with suitable higher coherences added
speaks about a generalization of (finite-dimensional) Hilbert spaces to higher ``$(\infty,1)$-Hilbert spaces'' (of finite type), such as modeled by (unbounded) $\RealNumbers$-chain complexes equipped with self-duality under the tensor product of chain complexes (cf. \cite[Ex. 2.4.28]{LurieTFT09}).
Progenitors of such $(\infty,1)$-Hilbert spaces appear in \cite{TQP}, where spaces of $\mathfrak{su}(2)$-anyonic quantum ground states arise as the homology  of chain complexes for twisted/equivariant cohomology of configuration spaces of points. Generally, topological quantum effects needed for topological quantum computing are described by (linear) higher homotopy theory (``higher structures'') of the kind expressed by the {\LHoTT} language.

\ifdefined\tentative
\newpage

On a more elemenary note, it may be worth highlighting that the semantics of \eqref{SelfDualRealModulesWithIsometricComplexStructure} in Real modules may readily be internalized into the formal language itself:

\noindent
{\bf The internal complex numbers.}
Using the negative unit scalar \eqref{TheNegativeUnitScalar}
we may define a commutative monoid internal to $\QuantumTypes$
\begin{equation}
  \label{InternalComplexNumbers}
  \ComplexNumbers
  \;\;
  \coloneqq
  \;\;
  \Big(
  \TensorUnit_{\Re}
  \!\oplus\!
  \TensorUnit_{\Im}
  ,\;
  (\mbox{-}) \cdot (\mbox{-})
  ,\;
  1
  \Big)
  \;\;
  \in
  \;\;
  \mathrm{CMon}\big(
    \QuantumTypes
    ,\,
    \otimes
    ,\,
    \TensorUnit
    ,\,
    \sigma
  \big)
  \,,
\end{equation}
(the subscripts just to keep track of the two copies of what in both cases is the tensor unit of $\QuantumTypes$)
with multiplication and unit given by the time-honored formulas for the complex numbers, sraightforwardly internalized into the monoidal category of module spectra:
\begin{equation}
  \label{InternalComplexMultiplication}
  \begin{tikzcd}[
    row sep=-1pt
  ]
  \ComplexNumbers
  \otimes
  \ComplexNumbers
  \ar[
    rr,
    "{ \cdot }"
  ]
  &&
  \ComplexNumbers
  \\[+6pt]
    \TensorUnit_{\Re}
    \!\otimes\!
    \TensorUnit_{\Re}
    \,\oplus\,
    \TensorUnit_{\Im}
    \!\otimes\!
    \TensorUnit_{\Im}
    \ar[
      rr,
      "{
        \mathrm{id}
        \;\oplus\;
        -\mathrm{id}
      }"{description}
    ]
    &&
    \TensorUnit_{\Re}
    \\
    \oplus
    &&
    \oplus
    \\
    \TensorUnit_{\Re}
    \!\otimes\!
    \TensorUnit_{\Im}
    \,\oplus\,
    \TensorUnit_{\Im}
    \!\otimes\!
    \TensorUnit_{\Re}
    \ar[
      rr,
      "{
        \mathrm{id}
        \;\oplus\;
        \mathrm{id}
      }"{description}
    ]
    &&
    \TensorUnit_{\Im}
    \mathrlap{\,,}
  \end{tikzcd}
  \hspace{1cm}
  \begin{tikzcd}[row sep=-1pt]
    \TensorUnit
    \ar[
      rr,
      "{ 1 }"
    ]
    &&
    \ComplexNumbers
    \\[+6pt]
    \TensorUnit
    \ar[
      rr,
      "{ \mathrm{id} }"
    ]
    &&
    \TensorUnit_{\Re}
    \\
    &&
    \otimes
    \\
    &&
    \TensorUnit_{\Im}
    \mathrlap{\,.}
  \end{tikzcd}
\end{equation}
Similarly, internal complex conjugation is the monoid endomorphism on $\ComplexNumbers$ given by
\begin{equation}
  \label{InternalComplexConjugation}
  \begin{tikzcd}[row sep=0pt]
    \ComplexNumbers
    \ar[
      rr,
      "{
        \overline{(\mbox{-})}
      }"
    ]
    &&
    \ComplexNumbers
    \\[7pt]
    \TensorUnit_\Re
    \ar[
      rr,
      "{ \mathrm{id} }"
    ]
    &&
    \TensorUnit_{\Re}
    \\
    \oplus
    &&
    \oplus
    \\
    \TensorUnit_\Im
    \ar[
      rr,
      "{
        -\mathrm{id}
      }"
    ]
    &&
    \TensorUnit_\Im
    \mathrlap{\,,}
  \end{tikzcd}
\end{equation}
which is an involutive equivalence, by \eqref{NegativeUnitScalar}.

In the model of paramaterized $H\RealNumbers$-module spectra by local systems of $\RealNumbers$-chain complex, this construction produces, up to equivalence, the dg-$\RealNumbers$-algebra which is concentrated on the $\RealNumbers$-algebra of complex numbers in degree 0.

The same construction works, of course,  in the $\infty$-topos of $R$-module spectra for any $E_\infty$-ring $R$. Over the sphere spectrum, $R \,\defneq\, \mathbb{S}$, it produces what might be called the ``homotopy complex integers''. The specification to  $R \,\defneq\, H\RealNumbers$ is eventually  necessary for actual application to quantum physics: Formal definition even of something as elementary as a Hadamard quantum gate requires that one can find a term $\sqrt{2}$ of type $\TensorUnit$.

\smallskip

\noindent
{\bf Real homotopy types.}
Left base change to the slice over $\mathbf{B}\ZTwo$ induces a monad $\possibly_{\ZTwo}$ whose modal types are equivalent to types in the slice
\[
  \begin{tikzcd}[
    row sep=10pt
  ]
    \ast
    \ar[
      rr,
      "{
        \exists !
      }"
    ]
    &&
    \mathbf{B}\ZTwo
    \\
    \Types
    \ar[out=30+180, in=-30+180,
      looseness=3.2,
      shorten=-2pt,
      shift right=2pt,
      "\scalebox{1.3}{${
          \mathclap{
            \possibly_{\ZTwo}
          }
          \hspace{4pt}
        }$}"{pos=.5, description}
    ]
    \ar[
      rr,
      shift left=14pt,
      "{
        \coprod_{\ZTwo}
      }"{description}
    ]
    \ar[
      from=rr,
    ]
    \ar[
      rr,
      shift right=14pt,
      lightgray,
      "{
        \prod_{\ZTwo}
      }"{description}
    ]
    \ar[
      rr,
      phantom,
      shift left=5pt,
      "{
        \scalebox{.7}{$\bot$}
      }"
    ]
    \ar[
      rr,
      phantom,
      shift right=5pt,
      "{
        \scalebox{.7}{$\bot$}
      }"{lightgray}
    ]
    &&
    \Types_{/\mathbf{B}\ZTwo}
    \mathrlap{\,,}
  \end{tikzcd}
  \hspace{.7cm}
  \Types_{\possibly_{\ZTwo}}
  \;\simeq\;
  \Types_{/\mathbf{B}\ZTwo}
  \,.
\]
This slice is still stable and distributive monoidal.

\smallskip

\noindent
{\bf The Real internal numbers.}
Equipped with the involution \eqref{InternalComplexConjugation},
the internal complex numbers \eqref{InternalComplexNumbers} become a monoid internal to $\possibly_\ZTwo$-modal type, as in the previous discussion:
\begin{equation}
  \RRealNumbers
  \;\;
  \defneq
  \;\;
  \ABundleType
    {
      \begin{tikzcd}
        \ComplexNumbers
        \sqcup
        \ComplexNumbers
        \ar[
          <->,
          bend right=25pt,
          shift right=10pt,
          start anchor={[xshift=-10pt]},
          end anchor={[xshift=+10pt]},
          "{ \overline{(\mbox{-})} }"{description}
        ]
      \end{tikzcd}
    }
    {
      \begin{tikzcd}
        \ast
        \sqcup
        \ast
        \ar[
          <->,
          bend right=20pt,
          shift right=8pt,
          start anchor={[xshift=-9pt]},
          end anchor={[xshift=+9pt]}
        ]
      \end{tikzcd}
    }
  \;\;
  \in
  \;\;
  \mathrm{Mon}\big(
    \Types_{\possibly_{\ZTwo}}
  \big)
  \,.
\end{equation}
We may think of this equivalently as monad structure (the $\RRealNumbers$-Writer monad) on the endofunctor that tensors with the underlying $\possibly_\ZTwo$-modal type
$$
  \begin{tikzcd}[sep=0pt]
    \QuantumTypes_{\possibly_\ZTwo}
    \ar[
      rr,
      "{}"
    ]
    &&
    \QuantumTypes_{\possibly_\ZTwo}
    \\
    \ABundleType
      { \VectorSpace{V}_\bullet }
      {  \HomotopyQuotient{X}{\ZTwo}  }
    &\mapsto&
    \RRealNumbers
    \otimes_{{}_{\mathbf{B}\ZTwo}}
    \ABundleType
      { \VectorSpace{V}_\bullet }
      { \HomotopyQuotient{X}{\ZTwo} }
  \end{tikzcd}
$$

\noindent
{\bf The Real quantum types.}
This way, the real quantum types are the modal types
$$
    \QuantumTypes_{
      \possibly_\ZTwo,
      \scalebox{.7}{$\RRealNumbers$}
    }
$$

\medskip

\medskip

\section{Quantization for Real}

Beyond purely linear/quantum types, there is classical/quantum interaction (through quantum measurement and state preparation) which is described by linear types {\it dependent} on classical types.

It is in this parameterized context that Real modules become strictly more general than $\RealNumbers$-vector spaces, since Real modules may not just depent on plain sets, but generally on {\it Real sets}:

\medskip

\noindent
{\bf The Real sets.}
We write
\begin{equation}
  \label{CategoryOfRealSets}
  \RealSets
  \;\;\;
  \defneq
  \;\;\;
  \Actions{\ZTwo}(\Sets)
  \;\;
  \simeq
  \;\;
  \mathrm{Func}\big(
    \mathbf{B}\ZTwo
    ,\,
    \Sets
  \big)
\end{equation}
 for the category of sets equipped with $\ZTwo$-actions (often ``$\ZTwo$-sets'', the {\it Real sets} among Atiyah's {\it Real spaces} \cite{Atiyah66}).

 There are two distinct embeddings of plain sets into Real sets:
\begin{itemize}[leftmargin=1cm]
  \item[{\bf (i)}]
  as sets with {\it trivial} action
  \begin{equation}
    \begin{tikzcd}[sep=0pt]
      \Set
      \ar[
        rr,
        hook,
        "{
        }"
      ]
      &&
      \RealSets
      \\
       W
        &\mapsto&
       \term{
         (\acts \, \ast)
         \times
         W
       }
    \end{tikzcd}
  \end{equation}
  This inclusion is fully faithful and reflects the ``discrete sets'' as seen internally in the topos of Real sets.
  \item[{\bf (ii)}]
  as sets with {\it free} action
  \begin{equation}
    \label{FormingFreeZTwoActionsOnSets}
    \begin{tikzcd}[sep=0pt]
      \Sets
      \ar[
        rr
      ]
      &&
      \RealSets
      \\
      W &\mapsto&
      (\ZTwo\!\acts \; \ZTwo)\times W
    \end{tikzcd}
  \end{equation}
  This inclusion is only faithful (though not far from full: the inclusion of hom-sets is ``of index two'') but is the left adjoint of a monadic functor:
\end{itemize}
\begin{equation}
  \label{BaseChangeToAndFromBZTwo}
  \begin{tikzcd}[
    row sep=20pt,
    column sep=55pt
  ]
    \ast
    \ar[
      rr,
      "{
        \vdash
        \,
        \mathrm{pt}
      }"{description}
    ]
    &&
    \mathbf{B}\ZTwo
    \ar[rr]
    &&
    \ast
    \\
    \Sets
    \ar[out=180+30, in=180-30,
      looseness=3.5,
      shift right=4pt,
      "\scalebox{1}{${
          \hspace{2pt}
          \mathclap{
            \smooth{(\mbox{-})}
          }
          \hspace{4pt}
        }$}"{pos=.5, description},
    ]
    \ar[
      rr,
      shift left=16pt,
      "{
        \ZTwo
        \!
        \acts \,
        (
        \ZTwo
        \times
        \mbox{-}
        )
      }"{description}
    ]
    \ar[
      from=rr,
      "{ U }"{description}
    ]
    \ar[
      rr,
      gray,
      shift right=16pt,
      "{
        \color{gray}
        \ZTwo
        \!
        \acts \,
        \mathrm{Hom}(
          \ZTwo
          ,\,
          \mbox{-}
        )
      }"{description}
    ]
    \ar[
      rr,
      phantom,
      shift left=7pt,
      "{
        \scalebox{.7}{$\bot$}
      }"
    ]
    \ar[
      rr,
      phantom,
      shift right=8pt,
      "{
        \scalebox{.7}{\color{gray}$\bot$}
      }"
    ]
    &&
    \RealSets
    \ar[
      rr,
      gray,
      shift left=16pt,
      "{
        \color{gray}
        (\mbox{-})/\ZTwo
      }"{description}
    ]
    \ar[
      from=rr,
      hook',
      "{
        (\ZTwo \!\acts \, \ast)
        \times (\mbox{-})
      }"{description}
    ]
    \ar[
      rr,
      shift right=16pt,
      "{
        (\mbox{-})^{\ZTwo}
      }"{description}
    ]
    \ar[
      rr,
      phantom,
      shift left=8pt,
      "{
        \scalebox{.7}{\color{gray}$\bot$}
      }"
    ]
    \ar[
      rr,
      phantom,
      shift right=8pt,
      "{
        \scalebox{.7}{$\bot$}
      }"
    ]
    &&
    \Sets
  \end{tikzcd}
\end{equation}

The category of Real sets \eqref{CategoryOfRealSets}
is equivalently the category of the corresponding action groupoids $\HomotopyQuotient{X}{\ZTwo}$ fibered over $\mathbf{B}\ZTwo \,\defneq\, \HomotopyQuotient{\ast}{\ZTwo}$:
\begin{equation}
  \begin{tikzcd}[sep=0pt]
    \RealSets
    \ar[
      rr,
      hook
    ]
    &&
    \Groupoids_{/\mathbf{B}\ZTwo}
    \\
    \ZTwo \acts \, X
    &\mapsto&
    \ABundleType
      {
        \HomotopyQuotient{X}{\ZTwo}
      }
      {
        \mathbf{B}\ZTwo
      }
  \end{tikzcd}
\end{equation}
This brings out the fact that we may think of Real sets as discrete $\ZTwo$-orbifolds (in fact: orientifolds, when we consider Real bundles over them):
The sets in the image of the free functor  correspond to the ``smooth'' (namely: non-singular) spaces among these orbifolds, which gives the name to the monad $\smooth{(-)}$ in \eqref{FormingFreeZTwoActionsOnSets}.

Hence Real sets are disjoint unions of smooth points $\smooth{\ast}$ and singular points $\HomotopyQuotient{\ast}{\ZTwo}$.

\medskip

\noindent
{\bf The Real bundles over Real sets.}
 The orbi/orienti-fold perspective \eqref{FormingFreeZTwoActionsOnSets} makes it manifest that the system of  categories
 of $\RealNumbers$-vector bundles over underlying sets equipped with $\RealNumbers$-vector bundle involutions covering those on the base is an ambidextrous pseudofunctor on Real Sets (for background on this and the following statements see \cite{EoS}):
 $$
   \begin{tikzcd}[
     row sep=0pt
   ]
   \RealSets
   \ar[
     rr
   ]
   &[-35pt]&[+35pt]
   \Categories_{\mathrm{adj}}
   \\
   \HomotopyQuotient{X}{\ZTwo}
   \ar[
     rr,
     phantom,
     "{ \mapsto }"
   ]
   \ar[dr]
   \ar[
     dd,
     "{ f }"{swap}
   ]
   &&
   \DependentModules
     { \RealNumbers }
     { \ARealSet{X} }
   \;\defneq\;
   \mathrm{Func}\big(
     \HomotopyQuotient{X}{\ZTwo}
     ,\,
     \Modules{\RealNumbers}
   \big)
   \ar[
     dd,
     shift right=65pt,
     "{ f_! }"{description}
   ]
   \ar[
     dd,
     shift right=58pt,
     phantom,
     "{
       \scalebox{.7}{$\dashv$}
     }"
   ]
   \ar[
     from=dd,
     shift left=50pt,
     "{
       f^\ast
     }"{description}
   ]
   \ar[
     dd,
     shift right=65-70pt,
     "{
       \mathrm{Lan}_f
     }"{swap}
   ]
   \ar[
     dd,
     shift right=58-70pt,
     phantom,
     "{
       \scalebox{.7}{$\dashv$}
     }"
   ]
   \ar[
     from=dd,
     shift left=50-70pt,
     "{
       \mathrm{Func}(f,\Modules{\RealNumbers})
     }"{swap}
   ]
   \\[10pt]
   &
   \mathbf{B}\ZTwo
   \\[10pt]
   \HomotopyQuotient{Y}{\ZTwo}
   \ar[
     rr,
     phantom,
     "{ \mapsto }"
   ]
   \ar[
     ur
   ]
   &&
   \DependentModules
     { \RealNumbers }
     { \ZTwo \!\acts  Y }
   \;\defneq\;
   \mathrm{Func}\big(
     \HomotopyQuotient{Y}{\ZTwo}
     ,\,
     \Modules{\RealNumbers}
   \big)
   \,.
   \end{tikzcd}
 $$
The resulting Grothendieck construction is the category of such bundles over varying Real base sets with vector bundle morphisms covering the base maps
$$
  \underset{
    \scalebox{.7}{$\ARealSet{X} $}
  }{\int}
  \DependentModules
    { \RealNumbers }
    { \ARealSet{X} }
  \hspace{.7cm}
    =
  \hspace{.7cm}
  \left\{
  \begin{tikzcd}
  [
    row sep=1pt,
    column sep=2pt
  ]
    &
  \ABundleType
    { \VectorSpace{V}_\bullet }
    { \HomotopyQuotient{X}{\ZTwo} }
  \ar[
    rrrrrr,
    shift left=13pt,
    "{ \phi_\bullet }"
  ]
  \ar[
    rrrrrr,
    shift right=13pt,
    "{ f }"
  ]
  &[-10pt]&[-10pt]&&&&[-10pt]
  \ABundleType
    { \VectorSpace{V}'_\bullet }
    { \HomotopyQuotient{X'}{\ZTwo} }
  &[-10pt]
  \\
  \VectorSpace{V}_{x}
  \ar[
    rrrrrr,
    "{
      \phi_x
    }"{description}
  ]
  \ar[
    <->,
    ddrr,
    "{
      \sim
    }"{sloped}
  ]
  &&&&&&
  \VectorSpace{V}'_{f(x)}
  \ar[
    <->,
    ddrr,
    "{ \sim }"{sloped}
  ]
  &
  \\
  \scalebox{.8}{$x$}
  \ar[
    <->,
    ddrr,
    shorten >=-2pt,
    "{ \sim }"{sloped}
  ]
  &&&&&&
  \\
  &&
  \VectorSpace{V}_{x'}
  \ar[
    rrrrrr,
    "{
      \phi_{x'}
    }"{description, pos=.6}
  ]
  &&&&&&
  \VectorSpace{V}'_{f(x')}
  \\
  &&
  \scalebox{.8}{$x'$}
  \end{tikzcd}
  \right\}
  \,.
$$
This becomes a symmetric monoidal category under the ``external'' tensor product $\externaltensor_{{}_\RealNumbers}$ of bundles
with tensor unit the $\RealNumbers$-line regarded as a bundle over the point equipped with the trivial $\ZTwo$-action.

The Real numbers constitute a monoid in this monoidal category
\[
  \RRealNumbers
  \;\;
  \defneq
  \;\;
  \ABundleType
    {
      \ComplexNumbers
        _{\scalebox{.6}{$\overline{(\mbox{-})}$}}
    }{
      \HomotopyQuotient
        { \ast }
        { \ZTwo }
    }
  \;\in\;
  \mathrm{Mon}
  \left(
    \underset{
      \scalebox{.7}{$\ARealSet{X}$}
    }{\int}
    \DependentModules
      { \RealNumbers }
      { \ARealSet{X} }
    ,\;
    \externaltensor_{{}_\RealNumbers}
    ,\;
    \ABundleType{
      \RealNumbers_{\mathrm{id}}
    }{
      \HomotopyQuotient{\ast}{\ZTwo}
    }
  \right)
\]
and its category of modules is equivalently Atiyah's Real vector bundles \cite{Atiyah66} over Real sets.
\begin{equation}
  \label{CategoryOfRealBundlesOverRealSets}
  \DependentModules
    { \RRealNumbers }
    { \RealSets }
  \;\;\;
  \defneq
  \;\;\;
  \Modules
    { \RRealNumbers }
  \left(
    \underset{
      \scalebox{.7}{$\ARealSet{X}$}
    }{\int}
    \DependentModules
      { \RealNumbers }
      { \ARealSet{X} }
    ,\;
    \externaltensor_{{}_\RealNumbers}
    ,\;
    \ABundleType{
      \RealNumbers_{\mathrm{id}}
    }{
      \HomotopyQuotient{\ast}{\ZTwo}
    }
  \right)
  \,.
\end{equation}
In particular, the Real modules over the point, from \eqref{CategoryOfRealModules}, are indeed the $\RRealNumbers$-modules over $\HomotopyQuotient{\ast}{\ZTwo}$ in the general sense of \eqref{CategoryOfRealBundlesOverRealSets}:
\[
  \Modules{\RRealNumbers}
  \;\;\simeq\;\;
  \DependentModules
    { \RRealNumbers }
    { \HomotopyQuotient{\ast}{\ZTwo} }
\]
because the $\RRealNumbers$-module structure implies the $\ComplexNumbers$-anti-linearity of the involution:
\[
  \begin{tikzcd}[
    column sep=-7pt, row sep=small
  ]
     \ComplexNumbers
    \ar[out=180-60, in=60,
      looseness=4.4,
      "\scalebox{1}{${
          \hspace{2pt}
          \mathclap{
            \overline{(\mbox{-})}
          }
          \hspace{4pt}
        }$}"{pos=.41, description},
        shift right=3pt
    ]
     &\otimes_{{}_\RealNumbers}\!\!&
     V
    \ar[out=180-60, in=60,
      looseness=4.4,
      "\scalebox{1}{${
          \hspace{2pt}
          \mathclap{
            \involution
          }
          \hspace{4pt}
        }$}"{pos=.41, description},
        shift right=3pt
    ]
    \ar[
      rrr
    ]
    &[10pt]
    &[10pt]
    &[0pt]
     V
    \ar[out=180-60, in=60,
      looseness=4.4,
      "\scalebox{1}{${
          \hspace{2pt}
          \mathclap{
            \involution
          }
          \hspace{4pt}
        }$}"{pos=.41, description},
        shift right=3pt
    ]
    \\[-9pt]
      z
      &
      \otimes
      \ar[
        dd,
        phantom,
        "{\mapsto}"{rotate=-90}
      ]
      &
      v
    &&\mapsto&
    z \cdot v
      \ar[
        d,
        phantom,
        "{\mapsto}"{rotate=-90}
      ]
    \\
    && && &
    \involution(z \cdot v)
    \\[-10pt]
    \ComplexConjugate{c}
    &\otimes&
    \involution(v)
    &\mapsto&
    \ComplexConjugate{c}
    \cdot
    \involution(v)
    \ar[
      ur,
      equals
    ]
    &
  \end{tikzcd}
\]

\medskip

\noindent
{\bf Complex vector bundles among Real vector bundles.}
Notice that the canonical ``embedding'' of $\ComplexNumbers$-vector bundles (over varying base sets) into Real vector bundles (over varying Real sets)
\begin{equation}
  \label{ComplexVectorBundlesAmongRealVectorBundles}
  \begin{array}{c}
  \begin{tikzcd}[
    column sep=13pt
  ]
    \DependentModules
      { \ComplexNumbers }
      { \Sets }
    \ar[
      rr
    ]
    &&
    \DependentModules
      { \RRealNumbers }
      { \RealSets }
  \end{tikzcd}
  \\
    \ABundleType
      { \VectorSpace{V} }
      { X }
   \;\;\;\mapsto\;\;\;
    \ABundleType
      {
        \begin{tikzcd}
        \VectorSpace{V}
        \sqcup
        \overline{\VectorSpace{V}}
        \ar[
          <->,
          shift right=5pt,
          bend right=23pt,
          start anchor={[xshift=-8pt]},
          end anchor={[xshift=+7pt]}
        ]
        \end{tikzcd}
      }
      {
        \begin{tikzcd}
        X
        \sqcup
        X
        \ar[
          <->,
          shift right=5pt,
          bend right=23pt,
          start anchor={[xshift=-8pt]},
          end anchor={[xshift=+7pt]}
        ]
        \end{tikzcd}
      }
  \end{array}
\end{equation}
is (faithful but) not full, cf. \eqref{FormingFreeZTwoActionsOnSets}: The morphisms on the right contain both (fiberwise) $\ComplexNumbers$-linear as well as $\ComplexNumbers$-anti-linear maps, depending on whether the base maps they cover are in the image of \eqref{ComplexVectorBundlesAmongRealVectorBundles} or compositions of these with the $\ZTwo$-action.

Instead, to get a full embedding one needs to consider, on the right, Real vector bundles equipped, internally, with  fiberwise complex structure, namely with endomorphisms that fix the base and square to $-1$ on the fibers. These act by multiplication with $\ImaginaryUnit$ on some fibers and with $-\ImaginaryUnit$ on their involution-mirrors, thus breaking the $\ZTwo$-symmetry. Specifically, the Real bundles which are images of $\ComplexNumbers$-linear bundles \eqref{ComplexVectorBundlesAmongRealVectorBundles} carry a canonical internal complex structure:
\begin{equation}
\label{InternalComplexStructureOnComplexBundleSeenAsRealBundle}
\begin{tikzcd}
  \VectorSpace{V}
  \sqcup
  \overline{\VectorSpace{V}}
  \ar[
    <->,
    shift right=5pt,
    bend right=23pt,
    start anchor={[xshift=-8pt]},
    end anchor={[xshift=+7pt]}
  ]
  \ar[
    rr,
    "{
      +\ImaginaryUnit
      \;\sqcup\;
      -\ImaginaryUnit
    }"
  ]
  \ar[
    d,
    ->>,
    shorten <=5pt
  ]
  &&
  \VectorSpace{V}
  \sqcup
  \overline{\VectorSpace{V}}
  \ar[
    <->,
    shift right=5pt,
    bend right=23pt,
    start anchor={[xshift=-8pt]},
    end anchor={[xshift=+7pt]}
  ]
  \ar[
    d,
    ->>,
    shorten <=5pt
  ]
  \\
  X
  \sqcup
  X
  \ar[
    <->,
    shift right=5pt,
    bend right=23pt,
    start anchor={[xshift=-8pt]},
    end anchor={[xshift=+7pt]}
  ]
  \ar[
    rr,
    equals
  ]
  &&
  X
  \sqcup
  X
  \mathrlap{\,.}
  \ar[
    <->,
    shift right=5pt,
    bend right=23pt,
    start anchor={[xshift=-8pt]},
    end anchor={[xshift=+7pt]}
  ]
\end{tikzcd}
\end{equation}


Interestingly, the $\ComplexNumbers$-line bundle over the point may be written as the Cartesian product of the smooth point with the Real numbers
\begin{equation}
  \label{ComplexLineAsSmoothPointTimesRealNumbers}
  \smooth{\ast} \times \RRealNumbers
  \;\;\;
  =
  \;\;\;
    \ABundleType
      {
        \begin{tikzcd}
        0
        \sqcup
        0
        \ar[
          <->,
          shift right=5pt,
          bend right=23pt,
          start anchor={[xshift=-8pt]},
          end anchor={[xshift=+7pt]}
        ]
        \end{tikzcd}
      }
      {
        \begin{tikzcd}
        \ast
        \sqcup
        \ast
        \ar[
          <->,
          shift right=5pt,
          bend right=23pt,
          start anchor={[xshift=-8pt]},
          end anchor={[xshift=+7pt]}
        ]
        \end{tikzcd}
      }
    \times
    \ABundleType
      { \RRealNumbers }
      { \ARealSet{\ast} }
    \;\;\;
    =
    \;\;\;
    \ABundleType
      {
        \begin{tikzcd}
        \ComplexNumbers
        \sqcup
        \overline{\ComplexNumbers}
        \ar[
          <->,
          shift right=5pt,
          bend right=23pt,
          start anchor={[xshift=-8pt]},
          end anchor={[xshift=+7pt]}
        ]
        \end{tikzcd}
      }
      {
        \begin{tikzcd}
        \ast
        \sqcup
        \ast
        \ar[
          <->,
          shift right=5pt,
          bend right=23pt,
          start anchor={[xshift=-8pt]},
          end anchor={[xshift=+7pt]}
        ]
        \end{tikzcd}
      }
\end{equation}

\medskip

\noindent
{\bf Real modules reflected inside Real bundles.} We may now observe that the canonical inclusion of Real modules into Real bundles is reflective, the reflector $\quantumly$ given by forming the direct sum of fibers equipped with the induced involution, we indicate the corresponding hom-isomorphism on the right:
\begin{equation}
  \label{ReflectionOfRealModulesInRealBundles}
  \begin{array}{c}
  \begin{tikzcd}
  \underset{
    \mathclap{
      x \in X
    }
  }{\oplus}
  \;
  \VectorSpace{V}_x
    \ar[out=56+90, in=-56+90,
      looseness=3,
      shorten=-3pt,
      "\scalebox{1}{${
        \VectorSpace{V}_x
        \leftrightarrow
        \VectorSpace{V}_{\sigma(x)}
        }$}"{pos=.5, description}
    ]
  \end{tikzcd}
  \hspace{-.2cm}
    \mapsfrom
  \hspace{.7cm}
  \ABundleType
    { \VectorSpace{V}_\bullet }
    {
      \HomotopyQuotient
        { X }
        { \ZTwo }
    }
  \\
  \begin{tikzcd}
    \Modules
      { \RRealNumbers }
    \ar[
      from=rr,
      ->>,
      shift right=7pt,
      "{ \quantumly }"{swap}
    ]
    \ar[
      rr,
      phantom,
      "{ \scalebox{.7}{$ \bot $} }"
    ]
    \ar[
      rr,
      shift right=7pt,
      hook
    ]
    &&
    \DependentModules
      { \RRealNumbers }
      { \RealSets }
  \end{tikzcd}
  \\
  \begin{tikzcd}
    \VectorSpace{V}
    \ar[out=40+90, in=-40+90,
      looseness=3.3,
      shorten=-3pt,
      "\scalebox{1}{${
          \hspace{3pt}
          \mathclap{
            C
          }
          \hspace{4pt}
        }$}"{pos=.5, description}
    ]
  \end{tikzcd}
  \hspace{.7cm}
  \mapsto
  \hspace{.9cm}
  \Bigg[
  \hspace{-6pt}
  \begin{tikzcd}[row sep=14pt]
    \VectorSpace{V}
    \ar[out=40+90, in=-40+90,
      looseness=3.3,
      shorten=-3pt,
      "\scalebox{1}{${
          \hspace{3pt}
          \mathclap{
            C
          }
          \hspace{4pt}
        }$}"{pos=.5, description}
    ]
    \ar[
      d,
      ->>
    ]
    \\
    \ast
    \ar[out=40-90, in=-40-90,
      looseness=3.3,
      shorten=-3pt,
      "\scalebox{1}{${
          \hspace{3pt}
          \mathclap{
            \ZTwo
          }
          \hspace{2pt}
        }$}"{pos=.5, description}
    ]
  \end{tikzcd}
  \hspace{-6pt}
  \Bigg]
  \end{array}
  \hspace{1.1cm}
  \begin{tikzcd}
  [
    row sep=1pt,
    column sep=2pt
  ]
    &
  \ABundleType
    { \VectorSpace{V}_\bullet }
    { \HomotopyQuotient{X}{\ZTwo} }
  \ar[
    rrrrrr,
    shift left=13pt,
    "{ \phi_\bullet }"
  ]
  \ar[
    rrrrrr,
    shift right=13pt,
    "{ \exists ! }"
  ]
  &[-25pt]&[-10pt]&&&&[-0pt]
  \ABundleType
    { \VectorSpace{V}' }
    { \mathbf{B}\ZTwo }
  &[-10pt]
  \\
  \VectorSpace{V}_{x}
  \ar[
    rrrrrr,
    "{
      \phi_x
    }"{description}
  ]
  \ar[
    <->,
    ddrr,
    "{
      \sim
    }"{sloped}
  ]
  &&&&&&
  \VectorSpace{V}'
  \ar[
    <->,
    ddrr,
    "{ \sim }"{sloped}
  ]
  &
  \\
  \scalebox{.8}{$x$}
  \ar[
    <->,
    ddrr,
    "{ \sim }",
    shorten >=-2pt
  ]
  &&&&&&
  \\
  &&
  \VectorSpace{V}_{x'}
  \ar[
    rrrrrr,
    "{
      \phi_{x'}
    }"{description, pos=.6}
  ]
  &&&&&&
  \VectorSpace{V}'
  \\
  &&
  \scalebox{.8}{$x'$}
  \end{tikzcd}
  \hspace{-.1cm}
  \leftrightarrow
  \hspace{-.1cm}
  \begin{tikzcd}
  \underset{
    \mathclap{
      x \in X
    }
  }{\oplus}
  \;
  \VectorSpace{V}_x
    \ar[out=56+90, in=-56+90,
      looseness=3,
      shorten=-3pt,
      "\scalebox{1}{${
        \VectorSpace{V}_x
        \leftrightarrow
        \VectorSpace{V}_{\sigma(x)}
        }$}"{pos=.5, description}
    ]
    \ar[
      rr,
      "{
        \underset{
          \mathclap{
            x \in X
          }
        }{\oplus}
        \;
        \phi_x
      }"
    ]
    &&
    \VectorSpace{V}'
    \mathrlap{\,.}
    \ar[out=30+90, in=-30+90,
      looseness=3,
      shorten=-3pt,
      "{ \sim }"
    ]
  \end{tikzcd}
\end{equation}

Notice that, by functoriality, this reflection sends complex structures on Real bundles, such as in \eqref{InternalComplexStructureOnComplexBundleSeenAsRealBundle}, to complex structures on Real modules, such as in \eqref{SelfDualRealModulesWithIsometricComplexStructure}.

\medskip

\noindent
{\bf The Real quantization.}
In enhancement of \cite{Monadology}, we now have a quantization (relative) monad $\quantized$ which produces not just linear spaces but Hermitian spaces, regarded as self-dual Real modules with isometric complex structure:

\begin{equation}
  \label{RealQuantization}
  \quantized \smooth{W}
  \;\;\;
  \defneq
  \;\;\;
  \quantumly
  \Big(
    \smooth{W}
    \times
    \RRealNumbers
  \Big)
\end{equation}

For finite $W$, the complex Real modules arising this way admit self-duality structure.

In particular the Real quantization of the non-singular point is the $\ComplexNumbers$-line
$$
  \quantized(\smooth{\ast})
  \;\;\;
  \underset{
    \mathclap{
    \scalebox{.7}{
      \color{gray}
      \eqref{RealQuantization}
    }
    }
  }{
    \simeq
  }
  \;\;\;
  \quantumly
  \Big(
    \smooth{\ast}
    \times
    \RRealNumbers
  \Big)
  \;\;\;
  \underset{
    \mathclap{
    \scalebox{.7}{
      \color{gray}
      \eqref{ComplexLineAsSmoothPointTimesRealNumbers}
    }
    }
  }{
    \simeq
  }
  \;\;\;
  \quantumly
    \ABundleType
      {
        \begin{tikzcd}
        \ComplexNumbers
        \sqcup
        \overline{\ComplexNumbers}
        \ar[
          <->,
          shift right=5pt,
          bend right=23pt,
          start anchor={[xshift=-8pt]},
          end anchor={[xshift=+7pt]}
        ]
        \end{tikzcd}
      }
      {
        \begin{tikzcd}
        \ast
        \sqcup
        \ast
        \ar[
          <->,
          shift right=5pt,
          bend right=23pt,
          start anchor={[xshift=-8pt]},
          end anchor={[xshift=+7pt]}
        ]
        \end{tikzcd}
      }
  \;\;\;
  \underset{
    \mathclap{
    \scalebox{.7}{
      \color{gray}
      \eqref{ReflectionOfRealModulesInRealBundles}
    }
    }
  }{
    \simeq
  }
  \;\;\;
        \begin{tikzcd}
        \ComplexNumbers
        \oplus
        \overline{\ComplexNumbers}
        \ar[
          <->,
          shift right=5pt,
          bend right=23pt,
          start anchor={[xshift=-8pt]},
          end anchor={[xshift=+7pt]}
        ]
        \end{tikzcd}
$$

This carries an essentially unique inner product structure, up to a sign. We can fix the sign by demanding that $\langle 1 \vert 1 \rangle = + 1$. That's the positivity condition.

\medskip

{\bf Remark.} The above is the Real version of the quantization modality in \cite{Monadology}. Notice that the analogous construction with $\RealNumbers$-bundles fails to provide complex spaces of quantum states. It is here that Real bundles are more than just an equivalent perspective on $\RealNumbers$-bundles.

\fi



\end{document}